\documentclass[12pt,english]{article}
\usepackage{amsmath,amssymb }
\usepackage{fancybox}
\usepackage{graphicx,psfrag,epsf}
\usepackage{enumerate}
\usepackage{graphicx,psfrag}
\usepackage{multirow}
\usepackage{epsfig}
\usepackage{subfigure}
\usepackage{theorem}
\usepackage{natbib,psfrag}
\usepackage[usenames,dvipsnames]{color}
\usepackage{todonotes}
\usepackage{xcolor}
\usepackage{algorithm2e}

\newcommand{\pdf}{0}
\if1\pdf
\usepackage[pdftex]{graphicx}
\else
\usepackage{graphicx}
\fi

\def\mb#1{\setbox0=\hbox{$#1$}
  \kern-.025em\copy0\kern-\wd0
  \kern.05em\copy0\kern-\wd0
  \kern-.025em\raise.0em\box0}

\makeatletter
\newcommand*\dashline{\rotatebox[origin=c]{90}{$\dabar@\dabar@\dabar@$}}
\makeatother

\newcommand\ind{\stackrel{\rm ind}{\sim}}

\newcommand{\argmax}[1]{\underset{#1}{\operatorname{arg}\,\operatorname{max}}\;}

\newcommand{\cov}{\text{Cov}}

\newcommand{\var}{\text{Var}}

\newcommand{\diag}{\text{diag}}

\newcommand{\logit}{\text{logit}}

\begin{document}

\baselineskip 1.8em
\parskip 1em

\title{Conditional Spectral Analysis of Replicated Multiple Time Series with Application to Nocturnal Physiology}


\author{Robert T. Krafty, Ori Rosen, David S. Stoffer, \\ Daniel J. Buysse, and  Martica H. Hall
\footnote{R. T. Krafty is Associate Professor, Department of Biostatistics, D. S. Stoffer is Professor, Department of Statistics, and D. J. Buysse and M. H. Hall are Professors, Department of Psychiatry, University of Pittsburgh (rkrafty@pitt.edu, stoffer@pitt.edu, buyssedj@upmc.edu, hallmh@upmc.edu).  O. Rosen is Professor, Department of Mathematical Sciences, University of Texas at El Paso (ori@math.utep.edu).   This work was supported by NIH grants R01GM113243, P01AG020677, R01HL104607, G12MD007592 and RR024153, NSF grants DMS1506882, 2G12MD007592 and DMS1512188, and NSA grant  H98230-12-1-0246. The authors thank the Editor, Associate Editor and two referees for comments and insights that greatly improved the manuscript.}}

\maketitle

\newpage

\setlength{\baselineskip}{22pt}  

\begin{center}
\section*{Abstract}
\end{center}
This article considers the problem of analyzing associations between power spectra of multiple time series
and cross-sectional outcomes when data are observed from multiple subjects.
The motivating application comes from sleep medicine, where researchers are able to non-invasively record physiological time series signals during sleep.  The frequency patterns of these signals, which can be quantified through the power spectrum, contain interpretable information about biological processes.  An important problem in sleep research is drawing connections between power spectra of time series signals and clinical characteristics; these connections are key to understanding biological pathways through which sleep affects, and can be treated to improve, health.    Such analyses are challenging as they must overcome the complicated structure of a power spectrum from multiple time series as a complex positive-definite matrix-valued function.  This article proposes a new approach to such analyses based on a tensor-product spline model of Cholesky components of outcome-dependent power spectra.  The approach flexibly models power spectra as nonparametric functions of frequency and outcome while preserving geometric constraints.  Formulated in a fully Bayesian framework, a Whittle likelihood based Markov chain Monte Carlo (MCMC) algorithm is developed for automated model fitting and for conducting inference on associations between outcomes and spectral measures.  The method is used to analyze data from a study of sleep in older adults and uncovers new insights into how stress and arousal are connected to the amount of time one spends in bed.

\noindent%
KEY WORDS: Bayesian Analysis; Coherence; Heart Rate Variability; MCMC; Multivariate Time Series; Sleep; Smoothing Spline; Spectral Analysis; Tensor-Product ANOVA; Whittle Likelihood.

\bigskip

\section{Introduction}
Innovations in data collection and storage have led to an increase in the number of biomedical studies that record multiple time series signals and outcome measures in multiple subjects.  For many time series, including common signals such as blood pressure, heart rate and electroencephalography (EEG), frequency patterns that are quantified through the power spectrum contain important information about biological processes.  Consequently, studies whose goals are to understand how underlying biological mechanisms are connected to behavioral and clinical outcomes often require an analysis of associations between outcomes and  power spectra of multiple time series.

Our motivating application comes from  a sleep study whose goal is to better understand the pathways that connect sleep to health and functioning.   In the study, heart rate variability (HRV) is recorded in subjects during a night of sleep.   HRV is measured through the series of elapsed times between consecutive heart beats, and its power spectrum  provides indirect measures of psychological stress and physiological arousal \citep{hall2007}.  Upon awakening, subjects reported subjectively assessed sleep outcomes, such as the amount of time slept during the night, which are associated with many aspects of well-being \citep{buysse2014}. Understanding the association between the power spectrum of HRV during different sleep periods (i.e. beginning, middle and end of the night) and self-reported sleep outcomes is essential to understanding how stress connects sleep to health and, consequently, for guiding the use of treatments of poor sleep.

In the biomedical literature, a two-stage approach is typically used to analyze such data. In the first stage, power collapsed within pre-selected frequency bands is estimated individually for each time series \citep{eurotask, hall2004}.   A power spectrum is a function of frequency; power collapsed within a frequency band is an integral of the power spectrum over a range of frequencies, which converts the functional parameter into a scalar measure.  In the second stage,  classical statistical methods, such as ANOVA and linear regression, are used to evaluate associations between these band-collapsed spectral measures and outcomes.   Such an approach has three major drawbacks.  First, it is highly dependent on the frequency-band collapsed measures selected and there exists a hot debate as to which measures should be considered and/or how they should be interpreted \citep{burr2007}.  Ideally, an analysis of such data should provide  global measures that can be used to understand the entire system while also providing a means to conduct inference on any frequency band-collapsed measure of potential interest.  Second, estimated power is treated as if it were not an estimate but the true unknown parameter, leading to inaccurate inference.  Finally, band-specific frequency measures are estimated for each time series separately, inhibiting the evaluation of patterns across series.  For instance, in our motivating example, this two-stage approach does not examine how the coherence in HRV between the beginning and end of the night is connected to sleep outcomes.

In the statistics literature, a considerable amount of research has been conducted on methods for analyzing functional variables, a thorough review of which is given by  \cite{wang2016}.  Included in this body of work are methods for analyzing associations between power spectra and outcomes when one time series is observed per subject  \citep{stoffer2010, krafty2013hall}.
When one observes multiple time series per subject and interest lies in frequency patterns both  within each series and across different series, the problem becomes considerably more challenging.  This is the case in our motivating study, where we are interested not only in stress and arousal during particular periods of sleep, but also in their persistence and coherence across periods.  While the power spectrum from a single time series is a positive real-valued function of frequency,  the power spectrum from multiple time series is a positive-definite Hermitian matrix valued function of frequency.   An analysis of associations between power spectra from multiple time series and study outcomes must be able to flexibly model associations while preserving this positive-definite Hermitian structure.

Efficient nonparametric methods that preserve the positive-definite Hermitian structure of spectral matrices have been developed
for the simpler, classical problem of estimating the power spectrum of a multivariate time series from a single subject by modeling Cholesky components of spectral matrices as functions of frequency  \citep{dai2004, rosen2007, krafty2013collinge}.
In this article, we extend this framework to develop a new approach to analyzing data from multiple subjects that models Cholesky components as functions of both frequency and outcome.  Rather than being curves as functions of frequency, components of spectral matrices under the proposed model are surfaces.  Changes in these surfaces with respect to the outcome provide nonparametric measures of association between outcomes and power spectra.  The proposed method is formulated in a fully Bayesian framework;  a MCMC algorithm based on the Whittle likelihood, or the asymptotic likelihood derived from the Fourier transform of the data, is developed for model fitting and inference.  The method allows one to evaluate the entire outcome-dependent power spectrum and to conduct nonparametric inference on the association between the outcome and any function of the power spectrum.

The rest of the article is organized as follows. Our motivating application, the AgeWise Sleep Study, is discussed in Section \ref{sec:agewise}.  A review of spectral analysis in the classical setting, where data are observed from a single subject, is given in Section \ref{babysteps}. The proposed methodology for analyzing time series from multiple subjects is presented  in Section \ref{sec:method}. The proposed method is used to analyze data from the motivating application in Section \ref{sec:application} and some final remarks are offered in Section \ref{sec:discussion}.

\section{The AgeWise Sleep Study}\label{sec:agewise}
An estimated 43\% of older adults report problems initiating or maintaining sleep \citep{foley1995}.  Poor sleep in older adults has been linked to  depression, heart disease, obesity, arthritis, diabetes  and stroke \citep{foley2004}.  With medical and scientific advances leading to an increase in the world's elderly population, the consequences of poor sleep in older adults pose a major public health concern.  The AgeWise study is a NIH-funded Program Project conducted at the University of Pittsburgh that seeks a better understanding of causes, effects, and treatments of poor sleep in older adults.  Towards this goal, we consider  $N=108$ men and women between  69--89 years of age who were observed during a night of in-home sleep.  Two types of data were collected in each subject.  First, subjects were observed during the night through ambulatory polysomnography (PSG), or the continuous collection of electrophysiological changes that occur during sleep.  Second, upon awakening, subjects completed the Pittsburgh Sleep Diary \citep{monk1994} to record self-reported sleep outcomes during the night.

As previously discussed, HRV is the series of elapsed times between heart beats.  It is of interest to researchers, as it reflects neurological control of the heart, and through this capacity, its power spectrum provides indirect measures of stress reactivity and arousal.
The PSG used in the study included an electrocardiograph (ECG) to monitor heart activity.  The ECG was used to locate the timing of heart beats, which were then differenced, detrended, cubic spline interpolated, and resampled at 1 Hz to compute HRV series throughout the night.

\begin{figure} 
\includegraphics[width=6.5in]{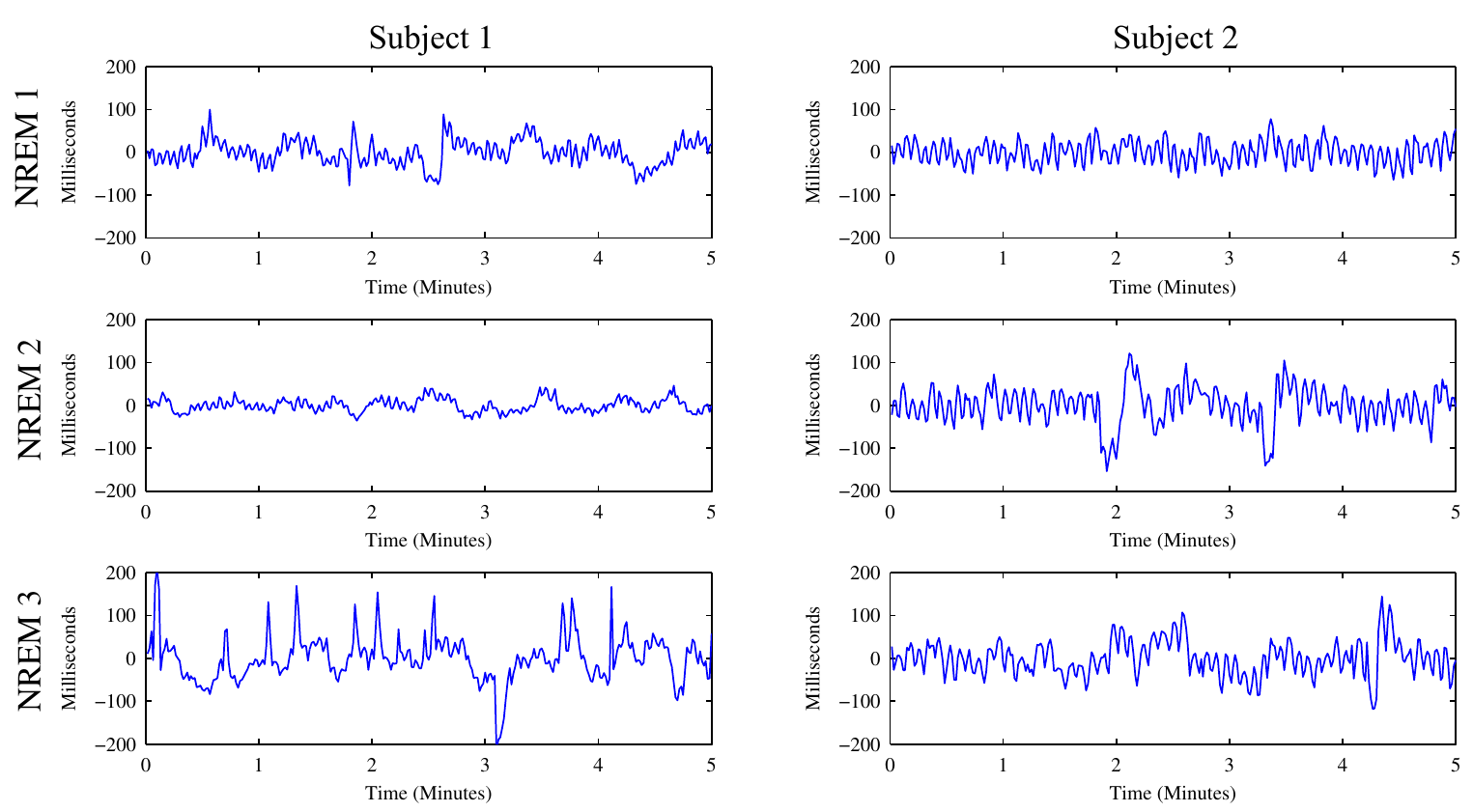}
 \caption{Detrended HRV time series during the first three periods of NREM from two subjects. Subject 1 reported a TIB of 357.67 minutes and subject 2 reported a TIB of 521.00 minutes.}
\label{fig:HRV}
\end{figure}

During the night, the body cycles through two types of sleep:  rapid eye movement (REM) and non-rapid eye movement (NREM) sleep.  In NREM sleep, which contains deep-sleep, the parasympathetic branch of the autonomic nervous system that is responsible for unconscious actions and stimulates the body to ``rest-and-digest'' dominates the sympathetic branch, which drives the ``flight-or-fight'' response.  Parasympathetic nervous system activity during NREM is hypothesized to be responsible for many of the rejuvenating properties of sleep \citep{siegel2005}.  However, physiological activity during NREM sleep is not constant; in good sleepers, the amount of parasympathetic activity during NREM increases throughout the night \citep{hall2004}.  To enable an analysis that can evaluate autonomic nervous system activity and its changes during the night,  we consider 3 HRV time series per subject (at the beginning, middle, and end of the night) by extracting the first 5 minutes of HRV from the first three periods of NREM sleep.  Data from two subjects are displayed in Figure \ref{fig:HRV}.

The goal of our analysis is to understand how the power spectrum of HRV over the three periods of NREM  are connected to self-reported sleep. We focus on one particular self-reported sleep measure derived from the Pittsburgh Sleep Diary:  time in bed (TIB).  TIB is defined as the elapsed time between attempted sleep and final wakening.  It serves as an upper bound for the amount of time spent asleep during the night, which has been linked to heart disease, hypertension, impaired neurobehavioral performance and mortality \citep{buysse2014}.  The reported TIB from our sample has a mean of 477.99 minutes
and a standard deviation of 71.32 minutes.
The resulting data for analysis consist of three epochs of HRV, one during each of the first three periods of NREM sleep, and self-reported TIB from each subject.

\section{Methodological Background: Spectral Domain Analysis}
\label{babysteps}

Before introducing our proposed model for the spectral analysis of multiple time series from multiple subjects in Section \ref{sec:method}, in this section we present background on spectral analysis in the classical setting, where data are observed from a single subject,  for both univariate and multivariate time series.

\subsection{Univariate Time Series}
\label{sec:uni}
\subsubsection{Population Parameters}
Spectral domain analysis focuses on the cyclic behavior of time series data.
An alternate approach is
 time domain analysis  wherein
the relationship between the data at different time lags is the focus.
For stationary time series, the main time domain tool is the covariance between
a current value of the series, say $X_t$, and the value of the series   $h$ time units prior, say
$X_{t-h}$.  The autocovariance is a function of lag, and is given by
\[\gamma(h)= \cov(X_{t},X_{t-h}), \quad h=0,\pm 1, \pm 2,\dots \,.\]
If $\gamma(h)$ is absolutely summable (which it is for ARMA models, for example), then
there is a duality between the power spectrum, given by
\begin{equation*}
f(\omega)=\sum_{h=-\infty}^\infty \gamma(h)\exp\left(-2\pi i\omega h\right),
\quad \omega \in \mathbb{R}, \label{baby1}\end{equation*}
and the autocovariance function, namely,
\begin{equation}
\gamma(h)=\int_{-1/2}^{1/2} f(\omega) \exp\left(2\pi i\omega h\right) d\omega, \quad
h =0, \pm1, \pm2, \ldots,
\label{eq:baby2}
\end{equation}
  as the inverse transform of  the
 power spectrum.
 The relationship is the same as that of a characteristic function and a probability
 density. Consequently, the information contained in the power spectrum is equivalent
 to the information contained in the autocovariance function.
 If we are concerned with lagged behavior, working with $\gamma(h)$ is more informative;  if we are concerned with cyclic behavior, as is the case of HRV where cyclical behavior provides interpretable physiological information, working with $f(\omega)$ is more informative.

The power spectrum is nonnegative and we assume that it is positive, so that
  $f(\omega) > 0$
  for all $\omega$.  In addition to being positive, $f$ has two other restrictions as a function of frequency.  By the nature of the Fourier transform, it is periodic such that $f(\omega) = f(\omega + 2 \pi)$, and it is a Hermitian function, or an even function, where $f(\omega) = f(-\omega)$.  Consequently, $f(\omega)$ is usually displayed only for $\omega \in [0,1/2]$.

  Putting $h=0$ in \eqref{eq:baby2} yields
 \[\gamma(0) =\var(X_t) = \int_{-1/2}^{1/2} f(\omega)\, d\omega\,,\]
 which expresses the total variance of the time series
 as the integrated power spectrum.
 In particular, we may think of $f(\omega)\, d\omega$ as the approximate variance in
 the data attributed to frequencies in a small band of width $d\omega$ around
 $\omega$.
 It is   common to view spectral analysis as an analysis of variance (ANOVA)
 of   time series data with respect to frequency.
 In fact, the power spectrum is a density of variance rather than of probability.

 \subsubsection{Estimation}\label{sec:uni:est}
 The nonparametric estimation of $f$ from an epoch of length $n$, $X_1, \dots, X_n$, can begin by considering the discrete Fourier transform (DFT)
 \[
 Y_{m}=n^{-1/2}\sum_{t=1}^n  X_{t}\exp(-2\pi i \omega_m t),
\]
where $\omega_m = m/n$ are the Fourier frequencies.
When $n$ is large, $Y_m$ are approximately independent mean-zero complex normal random variables with variances $f(\omega_m)$ for $m=1,\dots, M$,  $M= \lfloor (n-1)/2 \rfloor$,  and $Y_m = \overline{Y}_{n - m}$ \citep[Appendix C]{shumway2011}.   Consequently, the periodogram  $\left|Y_m\right|^2$ provides approximately unbiased but noisy estimates of $f(\omega_m)$.   Consistent estimates can be obtained by smoothing the periodogram across frequency using tools such as local averaging \citep[Chapter 4.5]{shumway2011}, splines \citep{pawitan1994}, and wavelets \citep{moulin1994}.

Our estimation approach for multiple time series from multiple subjects, which we develop in Section \ref{sec:method}, is based on Bayesian splines.  To motivate the development of the new methodology, in this subsection we discuss first a Bayesian smoothing spline model for univariate spectral analysis, then discuss a low-rank approximation.
Smoothing spline estimation balances the fit of a function to observed data with a roughness-based measure of regularity.  The Bayesian formulation of smoothing splines was first discussed in the case of Gaussian observations by \cite{kimeldorf1970} and \cite{wahba1978}.  Under the Bayesian formulation, the likelihood provides a measure of fit to observed data, and regularity is imposed through a mean-zero Gaussian prior on the functional parameter, which induces a prior for the roughness of the function.

For spectrum estimation, the large sample distribution of $Y_m$ provides the Whittle likelihood   \citep{whittle1953, whittle1954}
\begin{equation*}
L(Y\mid f) \approx \prod_{m=1}^M  f^{-1}(\omega_m)\exp\bigl\{- f^{-1}(\omega_m)
\left| Y_{m} \right|^2\bigr\}.
\label{eq:unilike}
\end{equation*}
 We adopt generic notation  throughout this article where $Y$ will denote all DFT data. Since $f$ is a positive function, $\log f$ is modeled rather than $f$ itself to avoid constraints.
Although general measures of regularity can be considered, we focus on measuring the roughness of a function through its integrated squared second derivative,
\[ {\cal P}\left(\log f\right) = \int_{0}^{1/2} \left\{\left[\log f \right]''(\omega)\right\}^2 d\omega.\]
The specification of the prior distribution begins by decomposing $\log f$ into a linear part (which is in the null space of ${\cal P}$) and a nonlinear part.  To define the prior distribution for the nonlinear part, consider the reproducing kernel of the seminorm defined by ${\cal P}$
\begin{equation*}
\label{eq:J}
J(\omega_i,\omega_j) =  \int_0^{1/2} \left(\omega_i - \nu\right)_+ \left(\omega_j - \nu \right)_+ d \nu,
\end{equation*}
where  $\left(\nu\right)_+ = \max\left(\nu,0\right)$  \citep[Section 2.3.1]{gu2013}. For the Bayesian smoothing spline model, the prior distribution for the log-spectrum can be formulated as
\begin{equation*}
\log f\left(\omega\right) =  a_1 + a_2 \omega + \sum_{j=1}^M J(\omega, \omega_j) z_j
\label{eq:logf}
\end{equation*}
where  $\mb{z} = \left(z_1,\dots,z_M\right)'\sim  N(\mb 0, \tau^2 J^{-1})$ is independent of $ \mb{a} = \left(a_1, a_1 \right)' \sim N(0,\sigma^2_{\alpha} I_2)$,  $J = \left\{J(\omega_i , \omega_j )\right\}$ is the $M \times M$ matrix of $J$ evaluated at the Fourier frequencies and $I_2$ is the $2 \times 2$ identity matrix.  The reproducing property of the kernel $J$ provides a simple form for the roughness of the log-spectrum, ${\cal P}\left(\log f\right) = \mb{z}' J \mb{z}$ \citep[Chapter 2]{gu2013},
from which it can be seen that the prior distribution on the coefficients $\mb{z}$ induces a prior on the roughness of the spectrum where ${\cal P}\left(\log f\right) \sim \tau^2 \chi^2_M$, and $\chi^2_M$ denotes a chi-squared
random variable on $M$ degrees of freedom.  The smoothing parameter $\tau^2 > 0$ balances the smoothness of the estimator to its fit to the data such that, conditional on $\tau^2$,  the Bayes estimate (posterior median) approaches a linear function  as $\tau^2 \rightarrow 0$ and  interpolates $\left|Y_m\right|^2$ as $\tau ^2\rightarrow \infty$.

Two approaches may be taken for estimation and inference under the Bayesian model: empirical Bayes and fully Bayes.  In the empirical Bayes approach, a data driven method, such as generalized cross-validation (GCV) or generalized maximum likelihood (GML), is used to select the smoothing parameter $\tau^2$.  The log-spectrum is then estimated from the posterior distribution conditional on $\tau^2$, and its median is equivalent to the frequentist smoothing spline obtained by minimizing a penalized Whittle likelihood as $\sigma^2_{\alpha} \rightarrow \infty$ \citep{gu1992, qin2008}.  In the fully Bayesian approach, $\tau^2$ is treated as a random variable with a noninformative prior, and inference is conducted averaged over the posterior distribution of $\tau^2$  \citep{speckman2003, crainiceanu2005}.  Our proposed methodology will adopt the later approach, as discussed in Section \ref{sec:method}.

The presented smoothing spline model contains a large number of coefficients, which can impede computation and limit practicality.  Low rank approximations, such as those obtained by using a subset of the kernel functions $J(\cdot, \omega_j)$ \citep{gu2002kim} or another set of basis functions contained in the column space of $J$ \citep{wood2006}, can be used to ease computational burden without sacrificing model fit.  Here, we consider
the basis formed from the scaled eigenvectors of $J$, which has been used for power spectrum estimation by \cite{wood2002}, \cite{rosen2009} and \cite{rosen2012}.  This basis can model smooth functions with a relatively few number of basis
functions, provides a diagonal prior covariance structure that aids computation, and maintains the intuitive interpretation of the prior distribution regularizing roughness as measured through ${\cal P}$ \citep{nychka1996}.   To formulate this low-rank approximation, we first consider an equivalent formulation of the Bayesian smoothing spline model for $\log f$ at the Fourier frequencies.
Let $\log {\mb f} = \left[ \log f(\omega_1), \dots, \log f(\omega_M) \right]'$ be the log-spectrum evaluated at the Fourier frequencies.
 Further, let $J = V_J D_J V_J'$ be the spectral decomposition of $J$, $Q_J = V_J D_J^{1/2}$ and $\mb{c} = D_J^{1/2} V_J' \mb{z}$. Then an equivalent model at the Fourier frequencies is
\begin{equation*}
\log {\mb f} = L_J \mb{a} + Q_J \mb{c},
\end{equation*}
where $L_J = \left( \mb{1}_M \, \dashline  \, \mb{\omega} \right)$,  $\mb{1}_M$ is the $M$-vector of ones, $\mb{\omega} = \left(\omega_1, \dots, \omega_M\right)'$, and $\mb{c} \sim N(\mb{0}, \tau^2 I_M)$.

The eigenvectors contained in the columns of $V_J$ are in increasing order of roughness and the eigenvalues contained in the diagonal matrix $D_J$ decay rapidly \citep{eubank1999}.  Smooth functions can be accurately modeled through the first several $n_J < M$ to provide a low-rank approximation.
With a slight abuse of notation to avoid the need to introduce further variables, we let $Q_J$ represent the $M \times n_J$ matrix of the first $n_J$
eigenvectors with corresponding coefficients $\mb{c} \sim N\left(\mb{0}, \tau^2 I_{n_J} \right)$.

The selection of $n_J$ provides a compromise between low-rank computational feasibility and loss of flexibility relative to full $M$-rank model.  An intuitive measure of the loss of flexibility is the fraction of the total variance of the covariance matrix $J$ explained by its first $n_J$ eigenvectors (FVE), or the sum of its first $n_J$ eigenvalues divided by the total sum of all of its $M$ eigenvalues.   \cite{wood2002}, \cite{rosen2009} and \cite{rosen2012} suggest using $n_J=10$ basis functions, which, in each of their settings considered, equates to an FVE of 97.975\%.  Our empirical findings support this suggestion, and we recommend selecting $n_J$ to achieve a 97.975\% FVE.  Under this rule, $n_J = 7$ for $n=[15,18]$, $n_J=8$ for $n=[19, 22]$, $n_J=9$ for $n \in [23, 40]$ and $n_J=10$ for $n \in [41, 10^4]$.

\subsection{Multivariate Time Series}
\label{sec:mult}
\subsubsection{Population Parameters}
 The ideas presented in the univariate case generalize to the multivariate case wherein we observe a
 $P$-dimensional vector-valued
 time series, say $\{\mb X_t\}$.  Under stationarity, the autocovariance function is a
 $P\times P$ matrix given by
 \[\Gamma(h)= \cov(\mb X_{t}, \mb X_{t-h}), \quad h=0,\pm 1, \pm 2,\dots \,.\]
 If $\sum_h
 ||\Gamma(h)|| < \infty$, the spectral density matrix of the
  series $\mb X_t$ is given by
  \begin{equation*} \label{eq:fmult} f(\omega) =
 \sum_{h=-\infty}^{\infty} \Gamma(h) \exp\left(-2\pi i\omega h\right),
 \quad \omega \in \mathbb{R} \,
 \end{equation*}
 and the inverse relationship is
\[ \Gamma(h)=\int_{-1/2}^{1/2}  f(\omega)\ \exp\left(2\pi i\omega h\right)\ d\omega, \quad
 h =0, \pm1, \pm2, \ldots\,.\]
For each $\omega \in \mathbb{R}$, $f(\omega)$ is a $P\times P$  non-negative definite Hermitian matrix
  with the diagonal elements,
  $f_{pp}(\omega)$ for $p=1,\dots,P$, being the spectra of the individual components,
  and the off-diagonal elements, $f_{qp}(\omega)$ for $q\ne p=1,\dots,P$, being the
  cross-spectra. Throughout this article, we assume that  $f(\omega)$ is non-singular for all $\omega \in \mathbb{R}$. As in the univariate case, $f$ is a periodic and Hermitian function of frequency were, for matrix-valued functions, Hermitian as a function of frequency is defined as $f(\omega) = f^*(-\omega)$, and $f^*(\omega)$ is the complex conjugate of $f(\omega)$.

 An important example of the application of the cross-spectrum
 is to
 the problem of linearly predicting one of the component series, say $X_{qt}$, from
 another component, say $X_{pt}$.
 A measure of the
 strength of such a
 relationship is the squared  coherence
  function defined as
 \begin{equation*}
 \rho_{qp}^2(\omega)=\frac{\vert f_{qp}(\omega)\vert^2}{
 f_{qq}(\omega)f_{pp}(\omega)}\, .\label{baby3}
  \end{equation*}
 This is analogous to conventional squared
 correlation between two finite-variance random variables; e.g.,
 $0\le \rho_{qp}^2(\omega)\le 1$.
 This analogy motivates the interpretation of  squared coherence as the
 squared correlation between two time series at frequency $\omega$.
 These ideas extend in an obvious way to the concept of multiple coherence and
 partial coherence functions  obtained from the full spectral matrix
 in much the same way that multiple correlation and partial correlation can be
 obtained from a covariance matrix.  Full details of these results may be found in
 \citet[Chapters  4 \& 7]{shumway2011}.

\subsubsection{Estimation}\label{sec:mult:est}
In the multivariate setting, let
 \[
\mb Y_{m}=n^{-1/2}\sum_{t=1}^n \mb X_{t}\exp(-2\pi i \omega_m t)
\]
be the DFTs of the data.  In this case, the Whittle likelihood is
\[
L(Y\mid f) \approx \prod_{m=1}^M \bigl |f^{-1}(\omega_m )\bigr |\exp\bigl\{-\mb Y_{m}^* f^{-1}(\omega_m)
\mb Y_{m}\bigr\},
\]
and the periodogram $\mb Y_m \mb Y_m^*$ is an approximately unbiased but noisy estimate of $f(\omega_m)$, from which consistent estimates can be obtained by smoothing.

While in the univariate setting the spectrum is smoothed on the logarithmic scale to preserve positivity, Cholesky components of spectral matrices can be smoothed to preserve positive-definiteness in the multivariate setting \citep{dai2004, rosen2007, krafty2013collinge}.  The modified Cholesky decomposition assures that, for a spectral matrix $f(\omega)$, there exists a unique $P \times P$ lower triangular complex matrix $\Theta(\omega)$ with ones on the diagonal and a unique $P \times P$ positive diagonal matrix $\Psi(\omega)$ such that
\begin{equation*}
f^{-1}(\omega) = \Theta(\omega) \Psi^{-1}(\omega)\Theta^*(\omega).
\end{equation*}
There are $P^2$--Cholesky components to estimate:  $\Re\left\{\Theta_{k \ell}\right\}$ and  $\Im\left\{\Theta_{k \ell}\right\}$ for $k > \ell = 1,\dots,P-1$, and $\Psi^{-1}_{k k}$ for $k=1,\dots,P$.  Since the diagonal terms
$\Psi_{k k}^{-1}(\omega) > 0$, we model $\log \Psi_{k k}^{-1}$.  Letting ${\mb \theta}_{k \ell} = \left[ \Theta_{k \ell}(\omega_1), \dots, \Theta_{k \ell}(\omega_M) \right]'$ and $\log {\mb \psi}^{-1}_{k k} = \left[ \log \Psi^{-1}_{k k}(\omega_1), \dots, \log \Psi^{-1}_{k k}(\omega_M) \right]'$, we model:
\begin{eqnarray}\label{eq:real}
 \Re\left\{ {\mb \theta}_{k \ell} \right\} &=& L_J \mb{a}_{r k \ell} + Q_J \mb{c}_{r k \ell}, \quad k > \ell =1,\dots,P-1 \\
  \Im\left\{ {\mb \theta}_{k \ell} \right\} &=& L_J \mb{a}_{i k \ell} + Q_J \mb{c}_{i k \ell}, \quad k > \ell =1,\dots,P-1 \\
\log {\mb \psi}^{-1}_{k k} &=& L_J \mb{a}_{d k k} + Q_J \mb{c}_{d k k}, \quad k=1,\dots,P, \label{eq:diag}
\end{eqnarray}
where $\mb{c}_{r k \ell} \sim N(\mb 0, \tau_{r k \ell}^2 I_{n})$, $\mb{c}_{i k \ell} \sim N(\mb 0, \tau_{i k \ell}^2 I_{n})$, $\mb{c}_{d k k} \sim N(\mb 0, \tau_{d k k}^2 I_{n})$, $\mb a_{r k \ell } \sim N( \mb 0,\sigma^2_{\alpha} I_2)$, $\mb a_{i k \ell } \sim N(\mb 0,\sigma^2_{\alpha} I_2)$ and $\mb a_{d k k} \sim N( \mb 0, \sigma^2_{\alpha} I_2)$.
Throughout this article, $r$, $i$ and $d$ are used to denote coefficients for real components of $\Theta$, imaginary components of $\Theta$ and the logarithm of the diagonal components of
$\Psi^{-1}$, respectively.

\section{Methodology: Replicated Multiple Time Series}
\label{sec:method}
The primary question considered in this article is how to assess the association between the power spectrum of $P$-variate time series of length $n$, $\left\{\mb X_{j 1},\dots, \mb X_{j n}\right\}$, and real-valued static variables, $U_j$, observed from $j=1,\dots,N$ independent subjects.   In the motivating study, there are $N=108$ participants, $U_j$ is  self-reported TIB, and $\mb X_{j t}$ are time series of HRV during the first $P=3$ periods of NREM.
To address this question, we first introduce a new measure  in Section \ref{sec:notation}, the conditional power spectrum, which quantifies associations between power spectra and outcomes. Then, in Section \ref{sec:model}, we  develop a tensor product model for the conditional power spectrum that extends the Bayesian spline model of Cholesky components of a single multivariate time series to account for dependence on both frequency and outcome.

As previously mentioned, there are two approaches to conducting a Bayesian analysis with splines:  empirical Bayes and fully Bayesian.  Each approach has strengths and weaknesses.  In the empirical Bayes approach, smoothing parameters are estimated through a data-driven procedure.  Estimates conditional on smoothing parameters can be quickly computed through Fisher's scoring or Newton-Raphson and conditional inference on the modeled functions (Cholesky components in our setting) can be conducted through approximate ``Bayesian confidence intervals'' \citep{gu1992}. In the fully Bayesian approach, smoothing parameters $\tau^2$ are treated as random variables with noninformative priors and MCMC techniques are used to sample from the joint distribution of coefficients and smoothing parameters \citep{speckman2003, crainiceanu2005}.  The sample simulated from the posterior distribution using MCMC  provides a natural means of conducting inference on any function of the spectrum averaged over the distribution of the smoothing parameters, which accounts for uncertainty in the smoothing parameters when conducting inference.  As will be illustrated in Section \ref{sec:application}, inference on squared coherence, univariate spectra, and integral functions thereof (all of which are nonlinear functions of modeled Cholesky components)  are of direct scientific interest.  We develop the proposed methodology under a fully Bayesian framework, presenting prior distributions in Section \ref{sec:bayes} and the sampling scheme in Section \ref{sec:samp}.

\subsection{Conditional Power Spectrum}
\label{sec:notation}
Without loss of generality, we formulate the methodology assuming that $U_i$ is scaled to take values within $[0,1]$.
To quantify the association between the power spectrum of the time series $\mb{X}_{j t}$ and the outcome $U_{j}$, we define the conditional power spectrum
\begin{equation*}
f(\omega, u) = \sum_{h = -\infty}^{\infty} \cov\left(\mb X_{j t}, \mb X_{j, t + h} \mid U_j = u \right)e^{- 2 \pi i \omega h}, \quad \omega \in \mathbb{R}, u \in [0,1] .
\end{equation*}
As with the power spectrum of a single multivariate time series, the spectral matrices $f(\omega, u)$ are positive-definite $P \times P$ Hermitian matrices, and $f(\cdot,u)$ is a periodic and Hermitian function of frequency for fixed $u$.
In a traditional spectral analysis without a cross-sectional variable, spectral measures such as $f_{p q}$ and $\rho^2_{pq} = \left|f_{p q}\right|^2/\left(f_{p p} f_{q q}\right)$ are curves as functions of frequency.  In the conditional setting, these are surfaces as functions of both frequency and the variable $u$.   How these functions change with respect to $u$ provides information as to how spectral measures are associated with the variable.

\subsection{Bayesian Tensor-Product Model of Cholesky Components} \label{sec:model}
As in the classical setting discussed in Section \ref{sec:mult:est}, where a multivariate time series observed from a single subject, to preserve positive definiteness, we model the Cholesky components.  Let
\begin{equation*}
f^{-1}(\omega, u) = \Theta(\omega, u) \Psi^{-1}(\omega,u)\Theta^*(\omega, u),
\label{eq:cholesky}
\end{equation*}
be the modified Cholesky decomposition of the conditional power spectrum.  We use Bayesian tensor product models for the $P^2$--unique Cholesky components, which decompose the bivariate functions into products of univariate functions of $\omega$ and of $u$.

Bayesian  models for Cholesky components as functions of $\omega$ were discussed in Section \ref{sec:mult:est}.  Similarly,  a low-rank approximate Bayesian smoothing spline model for a function of outcomes at the observed values can be formulated.  Since the domain of the outcome values is $[0,1]$, as opposed the domain of the frequency values $[0, 1/2]$, we consider the kernel \[ H(u_i, u_j) = \int_0^{1} \left(u_i-v \right)_+ \left(u_j - v \right)_+ dv. \]
Letting $\mb{u} = \left(u_1, \dots, u_N\right)'$, a low-rank model for functions of the outcome evaluated at the observed values is:
\begin{equation}\label{eq:fout}
L_H \mb{a} + Q_H \mb{b},
\end{equation}
where $L_H = \left( \mb{1}_N \, \dashline  \, \mb{u} \right)$, $Q_H$ is the $N \times n_H$ matrix of the first $n_H$ columns of $V_H D_H^{1/2}$, $H = V_H D_H^{1/2} V_H'$ is the spectral decomposition of the $N \times N$ matrix $H = \left\{ H(u_i, u_j) \right\}$, and $\mb{a} \sim N(\mb{0}, \sigma^2_{\alpha} I_2)$ is independent of $\mb{b} \sim N(\mb{0}, \tau^2 I_{n_H})$.

To write the tensor-product model at the observed frequency-outcome points, concatenate components across frequency and outcome to define the $NM$-vectors
\[
\mb \theta_{k \ell} = \left[ \left\{ \Theta_{k \ell}(\omega_1, u_1), \dots, \Theta_{k \ell}(\omega_M, u_{1}) \right\}, \dots, \left\{ \Theta_{k \ell}(\omega_1, u_N), \dots, \Theta_{k \ell}(\omega_M, u_{N}) \right\} \right]'
\]
for $k > \ell = 1,\dots,P-1$. Similarly define $\log \mb \psi^{-1}_{k k}$
for $k=1,\dots,P$. The real and imaginary parts of $\mb \theta_{k \ell}$, and
$\log \mb \psi^{-1}_{k k}$ can then be expressed as tensor products of the spline models for functions of frequency (given in Equations (\ref{eq:real}) - (\ref{eq:diag})) and outcome (given in Equation (\ref{eq:fout}))
\[
\begin{array}{lll}
\Re\bigl\{\mb\theta_{k\ell}\bigr\} & = & \{L_H \otimes L_J\}\, \mb a_{rk\ell}
+\{Q_H \otimes L_J \}\, \mb b_{rk\ell} +
\left\{  L_H \otimes  Q_J \right\}\mb c_{rk\ell}
 +  \bigl\{ Q_H \otimes  Q_J\bigr\}\mb d_{rk\ell}\\
\Im\bigl\{\mb\theta_{k\ell}\bigr\} & = & \{L_H \otimes L_J\}\, \mb a_{ik\ell}
+\{Q_H \otimes  L_J \}\, \mb b_{ik\ell} +
\left\{  L_H \otimes  Q_J \right\}\mb c_{ik\ell}
 +  \bigl\{ Q_H \otimes  Q_J\bigr\}\mb d_{ik\ell}\\
\log\mb \psi_{kk}^{-1} & = & \{L_H \otimes  L_J\}\,\mb a_{dkk}  +
\{Q_H \otimes  L_J \}\,\mb b_{dkk}  +\left\{ L_H \otimes  Q_J \right\}\mb c_{dkk}
 +  \bigl\{Q_H\otimes  Q_J\bigr\}\mb d_{dkk}.
\end{array}
\]
This model decomposes conditional Cholesky components into combinations of products of univariate functions of frequency and univariate functions of outcome.  The parameters $\mb a$ are coefficients for functions that are products of linear functions of both $\omega$ and $u$, $\mb b$ are coefficients for functions that are products of linear functions of $\omega$ and nonlinear functions of $u$, $\mb c$ are coefficients for functions that are products of nonlinear functions of $\omega$ and linear functions of $u$, and $\mb d$ are coefficients for functions that are products of nonlinear functions of $\omega$ and of $u$.

\subsection{Prior Distributions}
\label{sec:bayes}
We define two types of prior distributions: prior distributions on coefficients conditional on smoothing parameters and prior distributions on smoothing parameters.  The tensor product model naturally enables the formulation of prior distributions that regularize its components as univariate functions of frequency and outcome. Letting $\mb \eta_{rk\ell}=(\mb a'_{rk\ell},\mb b'_{rk\ell}, \mb c'_{rk\ell}, \mb d'_{rk\ell})'$, $\mb \eta_{ik\ell}=(\mb a'_{ik\ell},\mb b'_{ik\ell}, \mb c'_{ik\ell}, \mb d'_{ik\ell})'$and  $\mb \eta_{dkk}=(\mb a'_{dkk},\mb b'_{dkk}, \mb c'_{dkk}, \mb d'_{dkk})'$, conditional on smoothing parameters, we assume the diagonal Gaussian smoothing priors
\begin{align}
\mb\eta_{rk\ell} \sim N(\mb 0, D_{rk\ell}) & \text{ where } D_{rk\ell}=\diag(\sigma_{\alpha}^2 \mb 1'_{4},\, \tau_{\beta rk\ell}^2 \mb 1'_{2n_H},\, \tau_{\gamma rk\ell}^2 \mb 1'_{2n_J},\,
\tau_{\delta rk\ell}^2 \mb 1'_{n_H \times n_J}), \nonumber \\
\mb\eta_{ik\ell} \sim N(\mb 0, D_{ik\ell}) & \text{ where } D_{ik\ell}=\diag(\sigma_{\alpha}^2 \mb 1'_{4},\, \tau_{\beta ik\ell}^2 \mb 1'_{2n_H},\, \tau_{\gamma ik\ell}^2 \mb 1'_{2n_J},\,
\tau_{\delta ik\ell}^2 \mb 1'_{n_H \times n_J}), \nonumber \\
\mb\eta_{dkk} \sim N(\mb 0, D_{dkk})& \text{ where }  D_{dkk}=\diag(\sigma_{\alpha}^2 \mb 1_{4}',\, \tau_{\beta dkk}^2 \mb 1'_{2 n_H},\, \tau_{\gamma dkk}^2 \mb 1'_{2 n_J},\,
\tau_{\delta dkk}^2 \mb 1'_{n_H \times n_J}) \nonumber
\end{align}
where $\mb 1_{n}$ is the $n$--vector of ones.

Prior distributions on the smoothing parameters are placed by assuming that
$\tau_{\beta rk\ell}$,
$\tau_{\gamma rk\ell}$,
$\tau_{\delta rk\ell}$,
$\tau_{\beta ik\ell}$,
$\tau_{\gamma ik\ell}$,
$\tau_{\delta ik\ell}$, $k>\ell=1,\ldots,P-1$,
$\tau_{\beta dkk}$,
$\tau_{\gamma dkk}$,
$\tau_{\delta dkk}$, $k=1,\ldots,P$, are independent Half-$t(\nu,G)$ random variables  with pdf
$p(x)\propto [1+(x/G)^2/\nu)]^{-(\nu+1)/2}$, $x>0$, where the hyperparameters $\nu$
and $G$ are assumed known \citep{gelman:2006}.
Computationally, it is convenient to utilize the following scale mixture representation
\citep{wand:etal:2012}: $(\tau^2\mid g) \sim IG(\nu/2,\nu/g)$, $g\sim IG(1/2,1/G^2)$,
where $IG(a,b)$, is the inverse Gamma distribution with pdf
$p(x)\propto x^{-(a+1)}\exp(-b/x)$, $x>0$.
The larger the value of $G$, the less informative the
prior, and we set $G$ to a large fixed value.  We found analyses to be insensitive to the choice of $G$, with $G=10$ and $G=10^5$ giving indistinguishable results in both simulations and in the analysis of the AgeWise data.
The hyperparmeter $\sigma^2_{\alpha}$, which is the prior variance of the coefficients of the linear terms, is assumed to be a known large value.  In our computations, $\sigma^2_{\alpha} = 10^2$ and $\sigma^2_{\alpha}=10^7$ gave indistinguishable results.

\subsection{Whittle Likelihood, Sample Scheme and Inference}\label{sec:samp}
Given observed time series, we define the DFT for the $j$th subject at frequency $\omega_m$ as
\[
\mb Y_{jm}=n^{-1/2}\sum_{t=1}^n \mb X_{jt}\exp(-2\pi i \omega_m t).
\]
For large $n$,  conditional on $u_j$, $\mb{Y}_{j m}$ are approximately independent mean-zero complex Gaussian random variables.  This provides the conditional Whittle likelihood
\[
L(Y\mid f) \approx \prod_{j=1}^N\prod_{m=1}^M \bigl |f^{-1}(\omega_m,u_j )\bigr |\exp\bigl\{-\mb Y_{jm}^* f^{-1}(\omega_m,u_j)
\mb Y_{jm}\bigr\}.
\]

There are $P^2\left[ \left(n_H + 2 \right)\left(n_J + 2\right) + 3\right]$ parameters in the model of $f$: $\left(n_H + 2 \right)\left(n_J + 2\right)$ regression coefficients and $3$ smoothing parameters for each of the $P^2$ Cholesky components. We develop a sampling scheme to sample from the joint posterior distribution of the coefficients ${\mb \eta}$'s and smoothing parameters $\tau^2$'s conditional on the DFT $Y$ and the observed outcomes $\mb u$. To aid in developing this sampling scheme, it is advantageous to consider a more compact notation by defining
\[
Q=
\left(
L_H \otimes  L_J \quad \dashline \quad Q_H \otimes  L_J
\quad \dashline  \quad L_H \otimes  Q_J \quad \dashline \quad  Q_H\otimes  Q_J
\right)\]
 so that
\[
\Re\bigl\{\mb\theta_{k\ell}\bigr\} = Q \mb\eta_{r k \ell},\quad \Im\bigl\{\mb\theta_{k\ell}\bigr\} = Q \mb\eta_{i k \ell} \;\;\mbox{and} \;\;
\log\mb \psi_{kk}^{-1} = Q \mb\eta_{d k k}.
\]

Each iteration of the sampling scheme consists of three steps.  First, the coefficients corresponding to the real and imaginary components of $\Theta$ (${\mb \eta}_{r k \ell} $ and ${\mb \eta}_{i k \ell}$) are sequentially sampled as Gaussian random variables from their conditional posterior distributions conditional on the current values of all other parameters.
In the second step, the coefficients corresponding to the diagonal
elements of $\Psi^{-1}$ (${\mb \eta}_{d k k}$) are drawn.
The log of the conditional posterior distribution of these coefficients is
given by
\begin{equation}
\log p\bigl(\mb\eta_{dkk} \mid \mb v_k, D_{dkk}\bigr) \stackrel{c}{=} \sum_{j=1}^N\sum_{m=1}^M
\Bigl\{\mb q'_{jm} \mb\eta_{dkk}-\exp(\mb q'_{jm}\mb\eta_{dkk} )v_{kjm}  \Bigr\}-\frac{1}{2}\mb\eta'_{dkk}D_{dkk}^{-1}
\mb\eta_{dkk},
\label{eq:eta_diag}
\end{equation}
where $\mb q'_{jm}$ is the row of $Q$ corresponding to the $j$th subject and $m$th frequency,
$\mb v_k$ is a vector with components $v_{kjm}$ depending on $Y$
and on other parameters held fixed (its exact form is given in Appendix B), and $\stackrel{c}{=}$ denotes equality up
to a constant. Since this is not a known distribution,
$\mb\eta_{dkk}$ are drawn in a Metropolis-Hastings step.
The last step samples smoothing parameters from their posterior distributions conditional on other parameters.  For ease of notation, in what follows
we describe the sampling scheme for the case $P=3$. Further details are given
in Appendix B. After initializing
all the parameters, the $s$th iteration,
$1\leq s \leq S$, of the Gibbs sampler consists of the following steps.
\begin{enumerate}
\item Sample the coefficients corresponding to $\Theta$: \\
\renewcommand{\arraystretch}{1.5}
\begin{enumerate}
\item
Draw
\[
\begin{array}{lll}
({\mb \eta}^{(s)}_{r21} \mid Y, \mb\psi_{11}^{-1 (s-1)},\mb\theta_{31}^{(s-1)}, D_{r21}^{(s-1)}) & \sim & N(\mb \mu_{r21}, \Sigma_{r21}) \\
({\mb \eta}^{(s)}_{i21} \mid Y, \mb\psi_{11}^{-1 (s-1)},\mb\theta_{31}^{(s-1)}, D_{i21}^{(s-1)}) & \sim & N(\mb \mu_{i21}, \Sigma_{i21}) \\
\end{array}
\]
and update $\mb\theta_{21}^{(s)}=Q \mb\eta_{r21}^{(s)}+
i Q \mb\eta_{i21}^{(s)}$.
\item
Draw
\[
\begin{array}{lll}
({\mb \eta}^{(s)}_{r31} \mid Y, \mb\psi_{11}^{-1 (s-1)},\mb\theta_{21}^{(s)},
 D_{r31}^{(s-1)}) & \sim & N(\mb \mu_{r31}, \Sigma_{r31}) \\
({\mb \eta}^{(s)}_{i31} \mid Y, \mb\psi_{11}^{-1 (s-1)},\mb\theta_{21}^{(s)},
 D_{i31}^{(s-1)}) & \sim & N(\mb \mu_{i31}, \Sigma_{i31}) \\
\end{array}
\]
and update $\mb\theta_{31}^{(s)}=Q \mb\eta_{r31}^{(s)}+
i Q \mb\eta_{i31}^{(s)}$.
\item
Draw
\[
\begin{array}{lll}
({\mb \eta}^{(s)}_{r32} \mid Y, \mb\psi_{22}^{-1 (s-1)}, D_{r32}^{(s-1)})
 & \sim & N(\mb \mu_{r32}, \Sigma_{r32}) \\
({\mb \eta}^{(s)}_{i32} \mid Y, \mb\psi_{22}^{-1 (s-1)}, D_{i32}^{(s-1)})
 & \sim &  N(\mb \mu_{i32}, \Sigma_{i32})
\end{array}
\]
\end{enumerate}
\renewcommand{\arraystretch}{1}
and update $\mb\theta_{32}^{(s)}=Q \mb\eta_{r32}^{(s)}+
i Q \mb\eta_{i32}^{(s)}$.

The exact forms of the conditional means and covariances, $\mu_{c k \ell}$ and $\Sigma_{c k \ell}$, $c=r,i$, are given in Appendix B.
\item Sample coefficients corresponding to $\Psi^{-1}$: \\
    \begin{algorithm}[H]
    \SetAlgoLined
     \For{ $k = 1,2,3$}{
     \begin{enumerate}
     \item
    Draw $\mb\eta_{dkk}^{(s)} \sim t_\nu(\hat{\mb\eta}_{dkk},
    \hat{\Sigma}_{dkk})$, where
    $\hat{\mb\eta}_{dkk}$ is the maximizer of (\ref{eq:eta_diag}) and $\hat{\Sigma}_{dkk}$ is \\ the inverse of the observed information matrix at  $\hat{\mb\eta}_{dkk}$.
    \item
    Compute
    \[
     r^{(s)}=\frac{p\bigl(\mb\eta_{dkk}^{(s)} \mid \mb v_k, D_{dkk}\bigr)
     f_T(\mb\eta_{dkk}^{(s-1)})}
     {p\bigl(\mb\eta_{dkk}^{(s-1)} \mid \mb v_k, D_{dkk}\bigr)
     f_T(\mb\eta_{dkk}^{(s)})},
     \]
     where $f_T$ is the density of the $t_\nu(\hat{\mb\eta}_{dkk},
    \hat{\Sigma}_{dkk})$ distribution.
    \item
    With probability $\min(1,r^{(s)})$ accept $\mb\eta_{dkk}^{(s)}$, otherwise
    $\mb\eta_{dkk}^{(s)}=\mb\eta_{dkk}^{(s-1)}$.
    \item
    Update $\mb\psi_{kk}^{-1 (s)}= \exp(Q \mb\eta_{dkk}^{(s)})$.
    \end{enumerate}

            }
     \end{algorithm}
\item Sample smoothing parameters: \\
    \begin{algorithm}[H]
    \SetAlgoLined
  \For{ $\ell = 1,2$}{
            \For{ $k = 2,3$}{

               Draw \begin{eqnarray*}
\tau_{\beta rkl}^{2 \, (s) } & \sim & IG((n_b+\nu)/2,\,\mb b_{rkl}^{\prime (s)} \mb b^{(s)}_{rkl}/2 +\nu/g_{\beta rkl}^{(s-1)}) \\
\tau_{\gamma rkl}^{2 \, (s) } & \sim & IG((n_c+\nu)/2,\,\mb c^{\prime (s)}_{rkl} \mb c^{(s)}_{rkl}/2 +\nu/g_{\gamma rkl}^{(s-1)})
\\
\tau_{\delta rkl}^{2 \, (s) } & \sim & IG((n_d+\nu)/2,\,\mb d^{\prime (s)}_{rkl} \mb d^{(s)}_{rkl}/2 +\nu/g_{\delta rkl}^{(s-1)}) \\
g_{\beta rkl}^{(s)} & \sim & IG((\nu+1)/2,\, \nu/\tau_{\beta rkl}^{2 \, (s) }+1/G^2) \\
g_{\gamma rkl}^{(s)} & \sim & IG((\nu+1)/2,\, \nu/\tau_{\gamma rkl}^{2 \, (s) }+1/G^2) \\
g_{\delta rkl}^{(s)} & \sim & IG((\nu+1)/2,\, \nu/\tau_{\delta rkl}^{2 \, (s) }+1/G^2).
\end{eqnarray*}
             }
            }
    \end{algorithm}
    The smoothing parameters for the imaginary and diagonal components are similarly drawn from inverse gamma distributions.
\end{enumerate}

\subsubsection*{Point Estimates and Credible Intervals}
The sample generated via MCMC methods provides a means of obtaining point estimates and credible intervals for any function of the spectrum averaged over the distribution of smoothing parameters through the sample mean and percentiles of the empirical distribution of the function evaluated at each iteration of the sampling algorithm.  For instance, a measure of interest in the analysis of HRV is the log-spectrum from the $p$th period of NREM, $\log f_{p p}$.  Consider
\begin{equation*}
f^{(s)}(\omega_m, u_j) = \left\{\Theta^{(s)}(\omega_m, u_j) \Psi^{-1 \, (s)}(\omega_m,u_j)\Theta^{* \, (s)}(\omega_m, u_j)\right\}^{-1}
\label{eq:sth_spec_mat}
\end{equation*}
as the estimated spectral matrix at the $s$th iteration corresponding to
$u_j$ and $\omega_m$
with $p$th diagonal element $f^{(s)}_{p p}(\omega_m, u_{j})$.
The matrix $\Theta^{(s)}(\omega_m, u_j)$ has $k\ell$th element
\begin{equation*}
\theta^{(s)}_{k\ell jm}=\mb q'_{jm} \mb\eta^{(s)}_{rk\ell}+ i\, \mb q'_{jm}\mb\eta^{(s)}_{ik\ell},\;\; k > \ell=1,\ldots,P-1,
\end{equation*}
and $\Psi^{-1 \, (s)}(\omega_m,u_j)$ has $kk$th element $\exp\left(\mb q'_{kk} \mb \eta^{(s)}_{d k k}\right).$   If $S$ iterations of the sampling algorithm are run with a burn-in of $S_0$, then an estimate of $\log f_{p p}(\omega_m, u_{\ell})$ can be computed as the mean of the
values $\left\{\log f^{(s)}_{p p}(\omega_m, u_{\ell}) \,;\, S_0 \le s \le S \right\},$ and a 95\% credible interval computed as their 2.5 and 97.5 empirical percentiles.

Scientific interest also lies in measures collapsed across frequency.  For example, as will be discussed in the following section, in the analysis of HRV, collapsed power within the high-frequency band (HF) between 0.15-0.40 Hz
\begin{equation}\label{eq:hf}
 f^{HF}_p(u) = \int_{.15}^{.40} f_{p p}(\omega,u) d\omega
\end{equation}
provides a measure of autonomic nervous system activity during the $p$th NREM period among people with a TIB of $u$.  Letting $f^{HF \, (s) }_p(u_{\ell}) = W^{-1} \sum_{.15 \le \omega_m \le .40} f_{p p}^{(s)}(\omega_m,u_{\ell})$ where $W$ is the number of Fourier frequencies within the HF band, an estimate of $f^{HF}_p(u_{\ell})$ is given by the mean of the values $\left\{f^{HF \, (s) }_p(u_{\ell}) \,;\, S_0 \le s \le S \right\}$, and a 95\% credible interval is given by their 2.5 and 97.5 empirical percentiles.

\section{Application to the AgeWise Study}\label{sec:application}
We used the proposed methodology to analyze the association between TIB and the power spectrum of the first three periods of NREM from $N=108$ AgeWise subjects, as described in Section~\ref{sec:agewise}.   The method was fit using $n_H = n_J =10$ basis functions, with hyperparameters $G=\sigma^2_{\alpha} = 10^5$, and for  $S=3000$ iterations of the MCMC algorithm with a burn-in of $S_0=500$.  Note that, in this example, there are a total of $P^2\left(n_H + 2\right)\left(n_J+2\right)=1296$ coefficients and $3P^2 = 27$ smoothing parameters. The average run time per iteration was 5.46 seconds with a standard deviation of 0.24 seconds using the program that is available on the journal's website in Matlab 2016b and macOS Sierra v10.12.1 on a 2.9 GHz Intel Core i7 processor with 16 GB RAM.

Although all desired analyses are obtained from one MCMC chain, to aid the biological and clinical discussion of the results, we present the analysis in three stages.   First, in Section \ref{sec:explore} we examine the estimated period-specific spectra and squared coherences as frequency-outcome surfaces.  In the subsequent two stages, we explore power collapsed within certain frequency bands as functions of TIB:  first for power within periods in Section \ref{sec:within}, then for coherence between periods in Section \ref{sec:cross}.  The results from these analyses provide new insights into biological underpinnings of spending too little or too much time in bed.  In particular, our analysis suggests that  (i) short TIB is connected to elevated stress and arousal within-periods of NREM  towards the end of the night and  (ii) long TIB is associated with a persistence in arousal in the beginning of the night.

\subsection{Analysis of the Conditional Spectrum}\label{sec:explore}
\begin{figure} 
\includegraphics[width=6.4in]{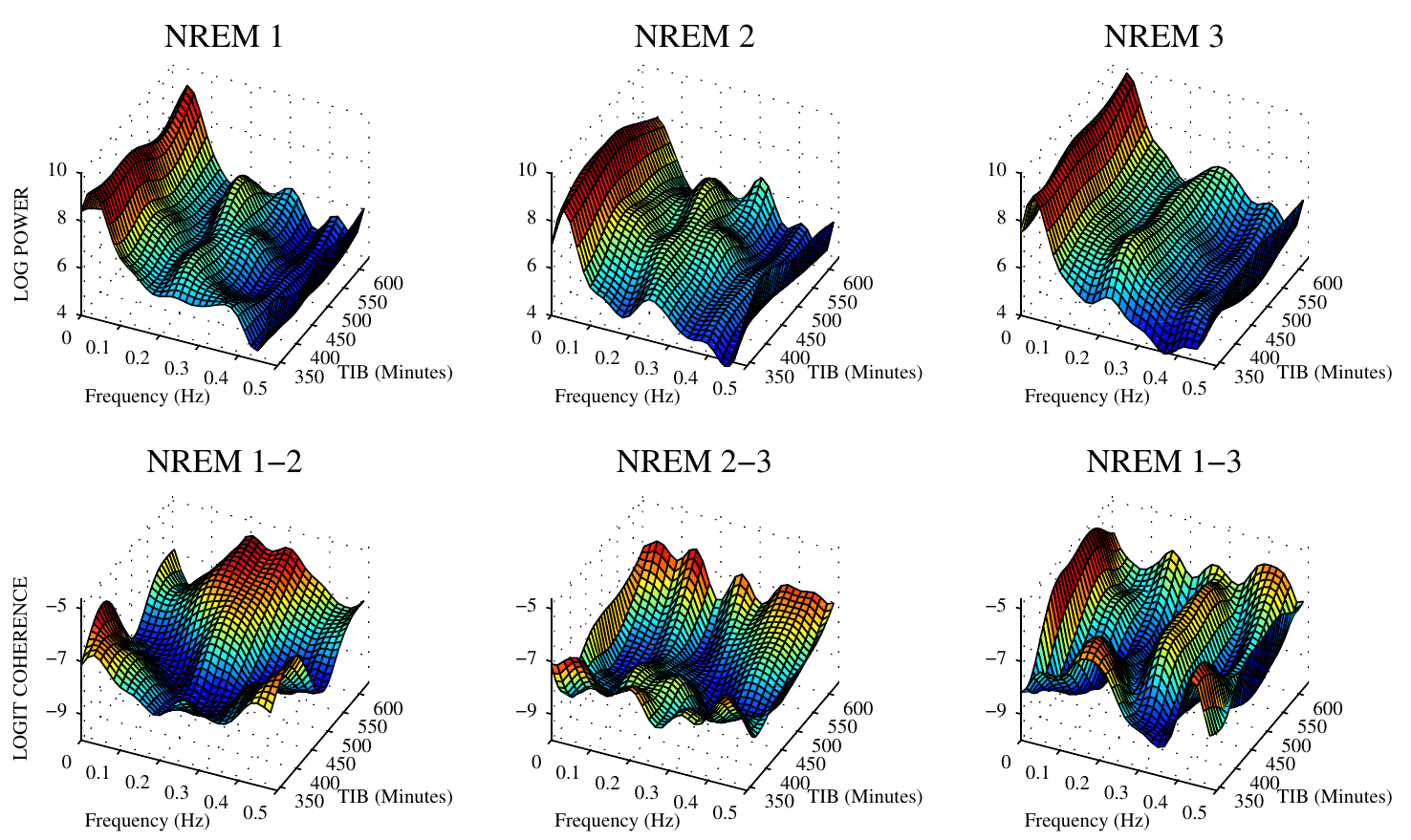}
 \caption{Estimated conditional log-spectra for each period of NREM (top panel) and estimated logit of conditional coherence between each period (bottom panel).}
\label{fig:SpecEst}
\end{figure}

Point estimates of the within-period conditional log-spectral surfaces, $\log\left\{f_{p p} (\omega, u)\right\}$, and of the cross-period conditional logit squared coherence surfaces,
\[\logit\left\{ \rho^2_{p q}(\omega, u) \right\} = \log\left[\rho^2_{pq}(\omega, u)/\left\{1-\rho^2_{pq}(\omega,u)\right\}\right],\] are displayed in Figure~\ref{fig:SpecEst}. These estimates are plotted on the logarithmic and logistic scales, respectively, to aid visualization.  The conditional spectra at each of the first three periods of NREM and the squared coherence between NREM 1 and 2 display different characteristics within low frequencies that are less than 0.15 Hz compared to higher frequencies between 0.15--0.40 Hz.

From a biological perspective, these results are not surprising and produce interpretable measures.  As was discussed in Section~\ref{sec:agewise}, the autonomic nervous system is classically divided into two branches:  the parasympathetic branch that is responsible for activities related to resting and digestion and the sympathetic branch that is responsible for the flight-or-fight response.  Researchers have shown that power within the high frequency band (HF) within 0.15-0.40 Hz provides a measure of parasympathetic nervous system activity and that power within the low frequency band (LF) between 0.04-0.15 Hz is a measure of the combined modulation of both the sympathetic and parasympathetic nervous systems.  Consequently, the ratio of power from low frequencies versus high frequencies (LF/HF) can be interpreted as a measure of sympathetic modulation relative to parasympathetic modulation. Blunted HF and elevated LF/HF power are often interpreted as indirect measures of physiological arousal and psychological stress  \citep{hall2004, hall2007}.  To obtain inference on associations between these measures and TIB, in the next two subsections we examine power and coherence collapsed within these bands as functions of TIB.

\subsection{Analysis of Within-Period Power}\label{sec:within}
\begin{figure} 
\includegraphics[width=6.5in]{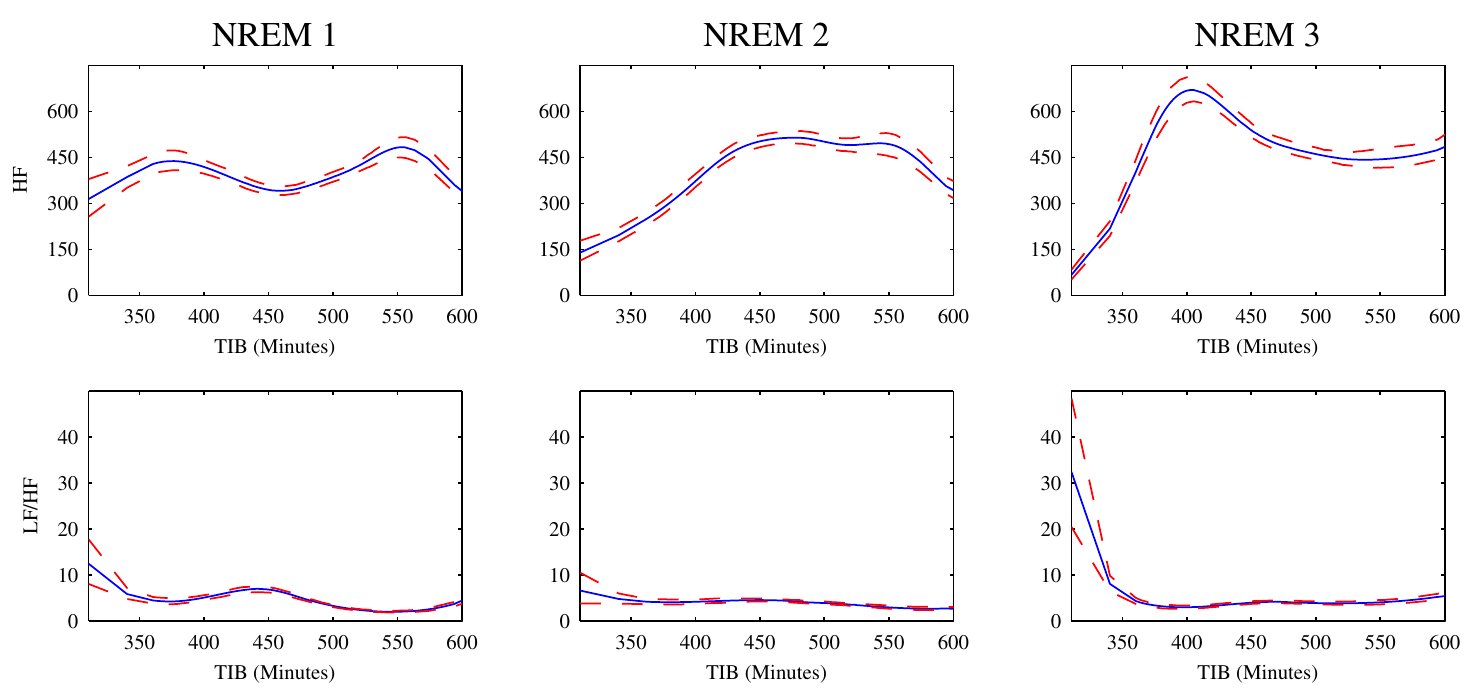}
 \caption{Estimated conditional HF  (top panel), $\hat{f}^{HF}_p$, and LF/HF (bottom panel), $\hat{f}^{LF/HF}_{p}$, as functions of TIB with 95\% pointwise credible intervals for each period of NREM.}
\label{fig:BandSpec}
\end{figure}

We consider two collapsed measures of within-period power.  In addition to HF previously defined in Equation (\ref{eq:hf}), we also consider LF/HF as
\begin{eqnarray*}
f^{LF/HF}_p(u) &=& \left\{\int_{.04}^{.15} f_{p p}(\omega,u) d\omega \right\} \bigg/ \left\{\int_{.15}^{.40} f_{p p}(\omega,u) d\omega \right\}.
\end{eqnarray*}
Estimates and 95\% pointwise credible intervals for these two measures as functions of TIB  are displayed in Figure \ref{fig:BandSpec} for each period.

HF power is relatively constant across  TIB  during NREM 1, while participants with a TIB of less than 400 minutes have decreased HF power during NREM 2 and 3 compared to those who spend more time in bed.  Further,  those who have an exceedingly small TIB  display increased LF/HF power during NREM sleep compared to those who spend more TIB, especially during NREM 3.  These characteristics are indicative of heightened physiological arousal and psychological stress.

Sleeping less than 7 hours per night has been shown to be associated with a multitude of negative health effects, including increased mortality \citep{buysse2014}.   The results of our analysis provide a potential pathway through which short sleep, which is inherently bounded by TIB, is connected to well-being: through increased stress and arousal towards the end of the night.

\subsection{Analysis of Cross-Period Coherence}\label{sec:cross}
\begin{figure} 
\includegraphics[width=6.5in]{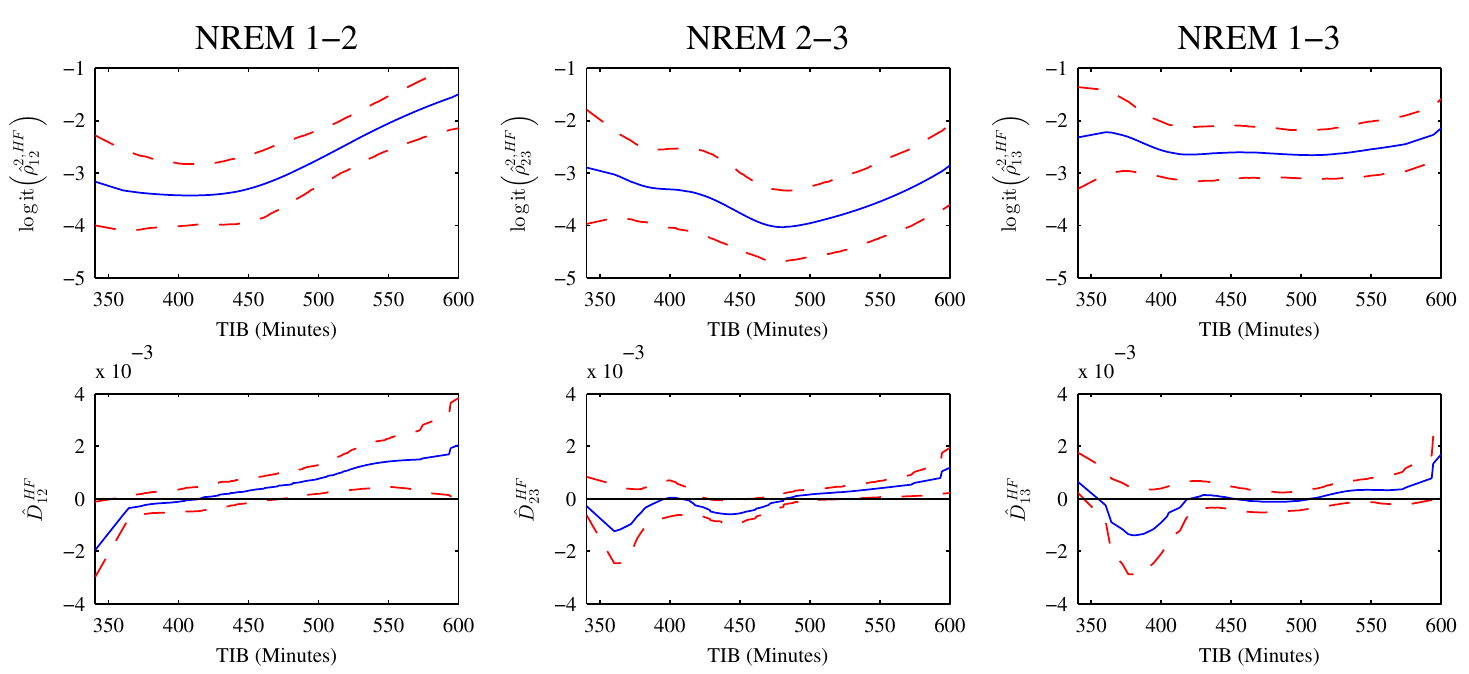}
 \caption{Estimated logit of integrated HF coherence (top panel), $\logit\left(\hat{\rho}^{2, \, HF}_{pq} \right)$, between each NREM period as functions of TIB and their first derivatives (bottom panel), $\hat{D}^{HF}_{pq}$, with pointwise 95\% credible intervals.}
\label{fig:BandCoher}
\end{figure}

To investigate  connections between cross-period coherence and TIB, we consider  conditional HF band-squared coherence
$$\rho^{2, \, HF}_{p q}(u) = \left|\int_{.15}^{.40} f_{p q}(\omega, u) d\omega\right|^2 \bigg/ \left\{f^{HF}_p(u) f^{HF}_q(u)\right\}$$
and display estimates on the logit scale, $\logit\left( \rho^{2,\,HF}_{pq}\right) = \log\left[ \rho^{2, \, HF}_{p q}/\left(1-\rho^{2, HF}_{p q}\right)\right]$, in the top panel of Figure \ref{fig:BandCoher}.   To better understand how changes in TIB are associated with HF coherence, we also examine first derivatives,  $$D^{HF}_{pq}(u) = d \left[\rho^{2, \, HF}_{pq}(u)\right]/d u,$$ whose estimates are displayed in the bottom panel of Figure \ref{fig:BandCoher}.  We find that the derivative of HF coherence between NREM 1 and 2 is positive for TIB greater than 500 minutes.  This indicates that excessive increases in the amount of time spent in bed are associated with increased coherence in parasympathetic activity in the beginning of the night.

The relationship between excessive TIB and ill-health led
\cite{youngstedt2004} to propose modest sleep restrictions to increase quality of life and survival, especially for older adults, who tend to spend more time in bed as compared to younger adults.  However, these restrictions must be used with great care as they can potentially lead to negative health effects \citep{reynolds2010, reynold2014}.  Our results demonstrate that excessive TIB is associated with a coherence in parasympathetic activity in the beginning of the night that is not present in moderate and short TIB.  A possible explanation for this relationship is that extensive TIB can cause an increase in the amount of time spent awake while in bed, or lead to fragmented sleep. The roles of and relationships between physiological activity during different sleep cycles could change as sleep becomes more fragmented.
These findings provide some of the first potential insights into the biological pathway through which excessive TIB can be connected to negative health, which can potentially be used to inform optimal sleep restriction strategies in older adults.

\section{Final Remarks} \label{sec:discussion}

This article introduces a novel approach to analyzing associations between multiple time series and cross-sectional outcomes when data are observed from multiple subjects.  A new measure of association, the conditional power spectrum, is introduced and its Cholesky components are modeled as bivariate functions of frequency and cross-sectional outcome.  A MCMC algorithm is developed for model fitting allowing for inference on any function of the power spectrum.  The method was motivated by  a sleep study and uncovered connections between excessive time in bed and heightened arousal and stress that could not have been uncovered through traditional methods.

We conclude this section by discussing three extensions to the proposed methodology.  First, the model is formulated to investigate the association between power spectra and a single cross-sectional variable.   The model could easily be extended through  higher-order tensor product models to include multiple variables, such as the amount of time it takes to fall asleep and the number of awakenings during the night.  However, such a model would provide inference on the effect of these variables on the power spectrum conditional on the other variables, complicating interpretation when these variables are highly correlated.
Future work will explore an interpretable canonical correlation type dimension reduction of a collection of correlated variables and multivariate spectral matrices, which can be viewed as a multivaraite extension of \cite{krafty2013hall}.  Second, our application focused on HRV,  due to the insights that it provides into autonomic nervous system activity.  One could also explore the spectral analysis of other PSG channels, as well as the simultaneous coupling of channels.   However, each channel of the PSG is sampled at a different rate.  The second extension will develop conditional spectral analysis of time series with different sampling rates.  Finally, since we were motivated by the analysis of HRV during epochs within NREM that are approximately stationary, we focused on stationary time series.
For more highly sampled signals such as EEG, this assumption is not valid.  A conditional time-frequency analysis for signals that are locally stationary will also be explored.

\section*{Code}

A Matlab program for implementing the proposed methodology and a file demonstrating its use are available at https://www.mathworks.com/matlabcentral/fileexchange/61186-mcbspec.


\bibliographystyle{asa}
\bibliography{covMultAbib}

\newpage

\section*{Appendices} 
\appendix
\gdef\thesection{\Alph{section}}
\numberwithin{equation}{section}
\numberwithin{figure}{section}
\numberwithin{table}{section}

\section{Simulation Study}
\begin{figure} 
\includegraphics[width=6.2in]{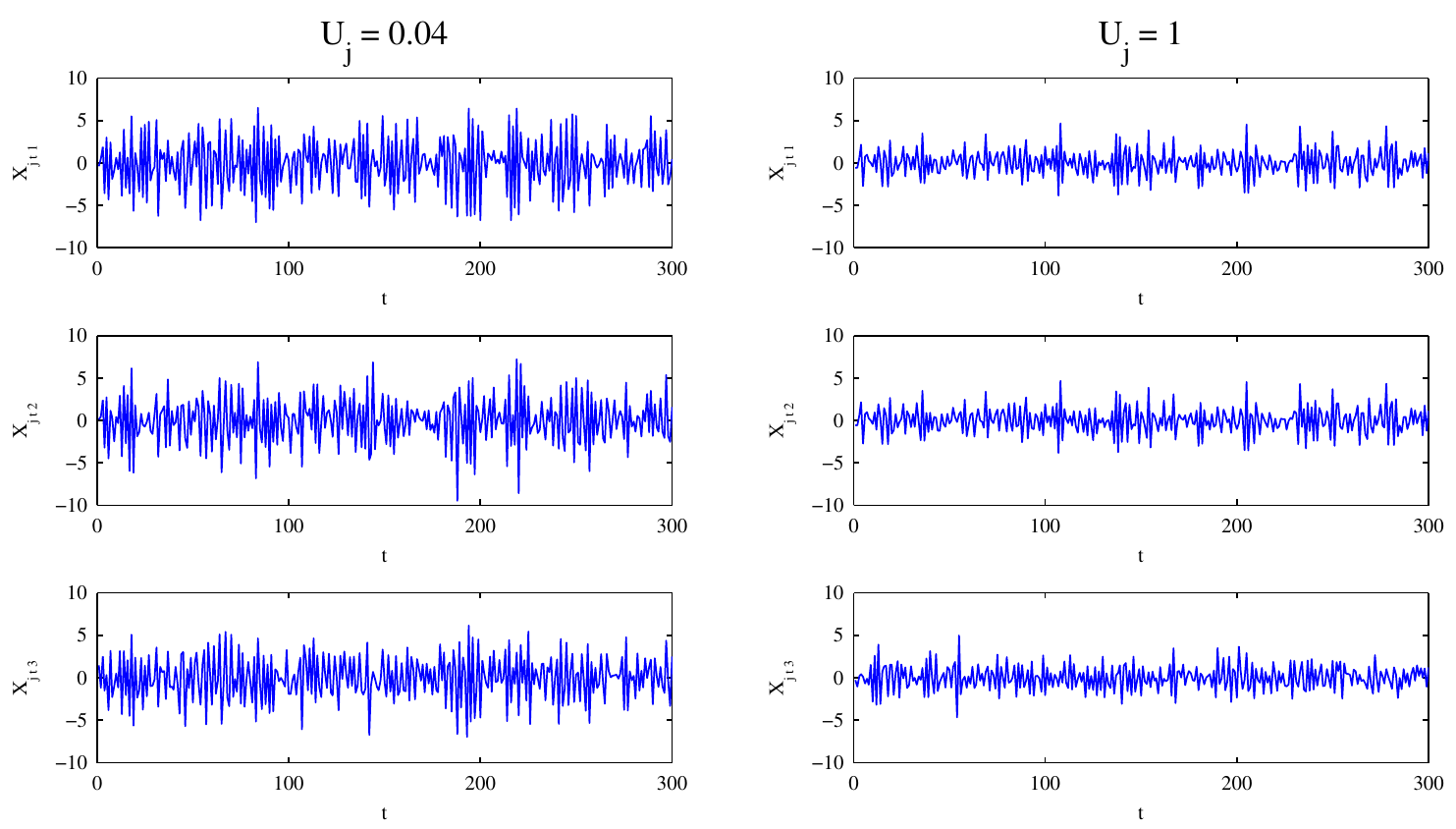}
 \caption{Simulated conditional MA(2) epochs of length $n=300$ with $u_j = 0.04$ and $u_j = 1$.}
\label{fig:tsplot}
\end{figure}

\begin{figure} 
\begin{center}
\includegraphics[width=5.9in]{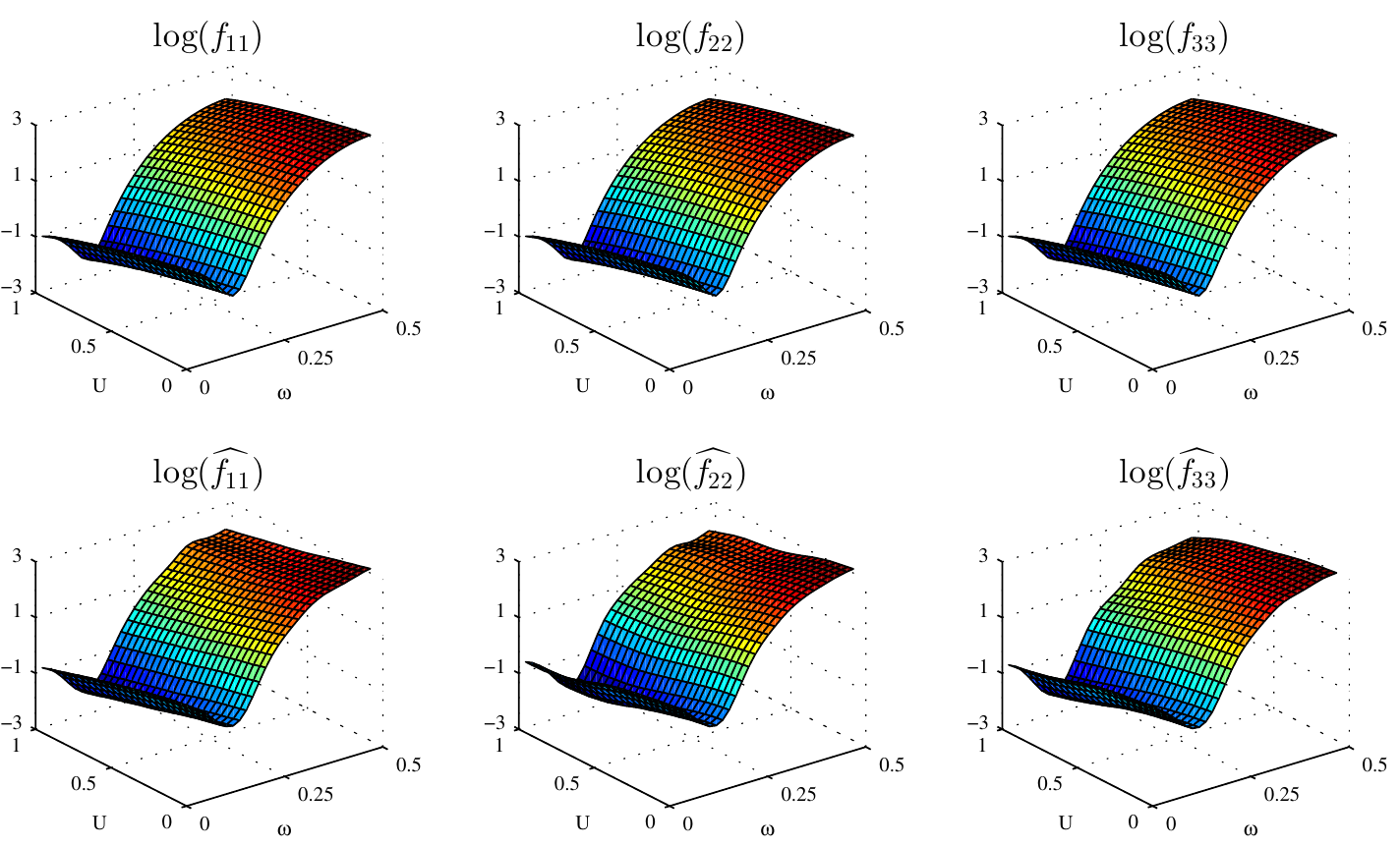}
\end{center}
 \caption{Conditional log-spectra from the MA(2) model (top panels) and estimated conditional log-spectra from a random sample of time series of length $n=300$ from $N=25$ subjects (lower panels). }
\label{fig:diag}
\end{figure}

\begin{figure} 
\begin{center}
\includegraphics[width=5.7in]{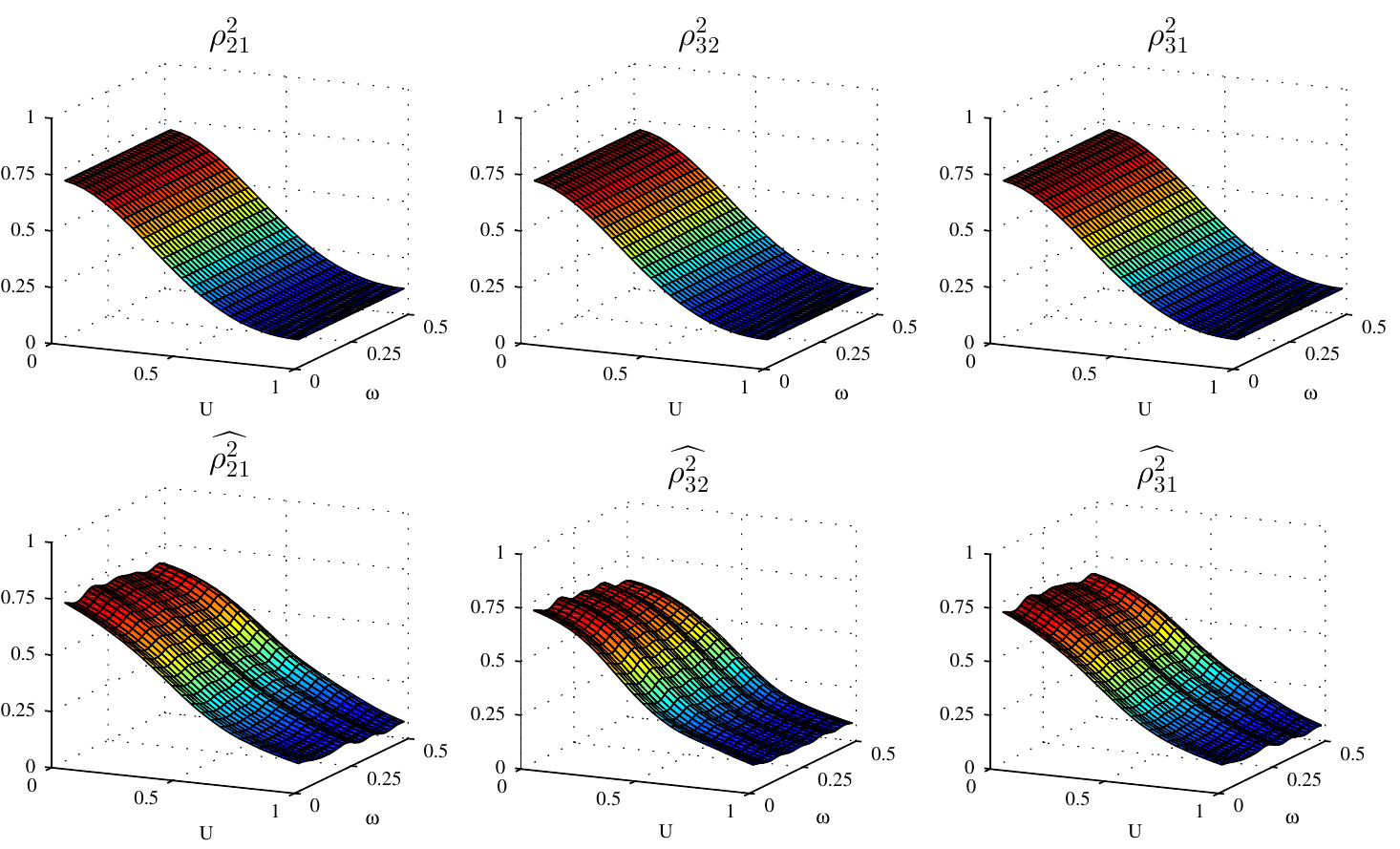}
\end{center}
 \caption{Conditional squared coherence from the MA(2) model (top panels) and estimated conditional squared coherence from a random sample of time series of length $n=300$ from $N=25$ subjects (lower panels). }
\label{fig:rho}
\end{figure}

\begin{figure} 
\begin{center}
\includegraphics[width=6in]{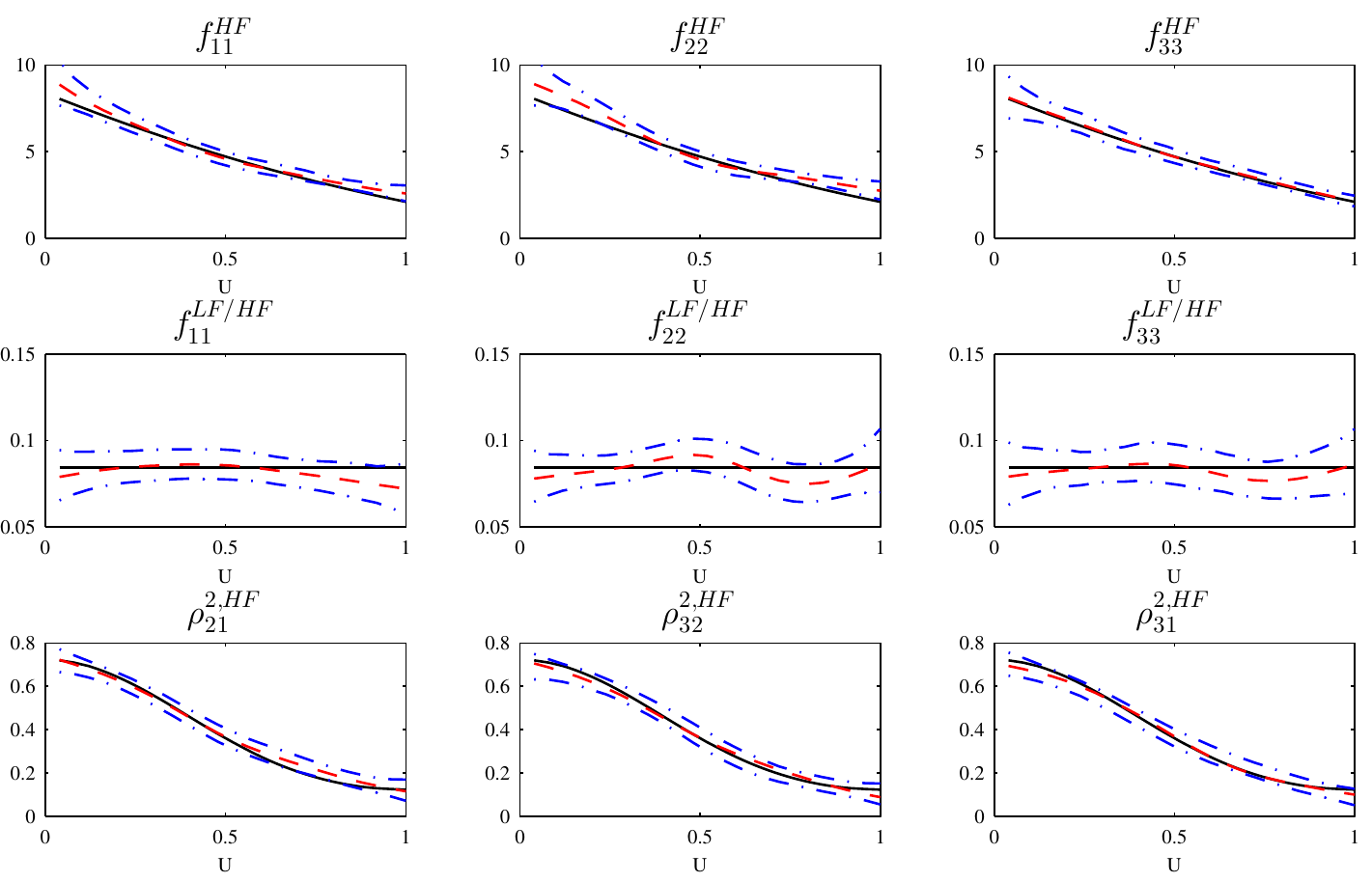}
\end{center}
 \caption{HF band power (top panels), LF/HF band power (middle panels), and HF band squared coherence (bottom panel) (---), along with point estimates  ({\color{red}- - -}) and 95\% credible intervals ({\color{blue}- $\cdot$ -}) from a random sample of $N=25$ conditional MA(2) time series of length $n=300$. }
\label{fig:band}
\end{figure}

To illustrate the proposed model and to investigate its empirical properties, we consider the $P=3$ dimensional second order moving average model ( MA(2) )
\[ \mb X_{j t} = \mb \epsilon_{j t} + \Theta_1 \mb \epsilon_{j \, t-1} + \Theta_2 \mb \epsilon_{j \, t-2}, \quad j=1,\dots,N, \quad t=1,\dots,n, \]
where $\Theta_1 = -  I$, $\Theta_2 = 0.6  I$, $I$ is the $3\times3$ identity matrix, and $\mb \epsilon_{j t}$ are independent $N\left[\mb 0, \Omega\left(u_j\right)\right]$ random variables  with
\[\Omega\left(u\right) = \sigma^2(u) \left[ \begin{tabular}{c c c}
                            1 & $\rho(u)$ & $\rho(u)$ \\
                            $\rho(u)$ & 1 & $\rho(u)$ \\
                            $\rho(u)$ & $\rho(u)$ & 1
                            \end{tabular}
                    \right],
                    \]
$\sigma^2(u) = \left(2 - u\right)^2$ and $\rho(u) = 0.6 + 0.25 \cos\left( \pi u\right)$.
Two simulated epochs of length $n=300$, one with $u_j = 0.04$ and one with $u_j = 1$, are displayed in Figure \ref{fig:tsplot}.
The conditional spectrum is given by
\[f(\omega, u) = \Theta(\omega) \Omega\left(u\right) \Theta(\omega)^*,\]
where $\Theta(\omega) =  I + \Theta_1 \exp(- 2 \pi i \omega) + \Theta_2 \exp(-4 \pi i \omega).$   The log-spectra, $\log\left[f_{p p}(\omega, u)\right]$, and their estimates under the proposed procedure from a random sample of $N=25$ independent epochs of length $n=300$  are displayed in Figure \ref{fig:diag}. Plots of the logit-squared coherence, $\text{logit}\left[\rho^2_{pq}(\omega, u)\right]$, and their estimates are displayed in Figure \ref{fig:rho}. The band-collapsed measures $f^{HF}_{p}(u)$, $f^{LF/HF}_{p}(u)$ and $\rho^{2, HF}_{p q}(u)$, along with their estimates and 95\% credible intervals, are displayed in Figure \ref{fig:band}.

We simulated 100 random samples of conditional MA(2) time series of length $n$ from $N$ subjects with $u_j = j/N$ for the four combinations of $n=300, 500$ and $N=25, 50$.  The estimation procedure was run using $n_J = 10$, $n_H= 5$, and for 2000 iterations of the MCMC algorithm with burn-in of 500 iterations.  To investigate the sensitivity of the proposed estimation procedure with respect to hyperparameters, the procedure was run twice:  for $G=10^5$ and for $G=10^{10}$. Table \ref{tab:RunTime} reports the mean and standard deviation of run times per iteration for each setting using the program that is available on the journal's website
in Matlab 2016b and macOS Sierra v10.12.1 on a 2.9 GHz Intel Core i7 processor with 16 GB RAM.

\begin{table}
    \centering
\renewcommand\arraystretch{.75}
\begin{tabular}{l | c c}
  & $n=300$& $n=500$  \\ \hline
$N=25$ & 0.35 & 0.92 \\
  & {\footnotesize {\em (0.02)}}   & {\footnotesize {\em (0.03)}} \\
  $N=50$ & 1.29 & 3.40 \\
  & {\footnotesize {\em (0.03)}}   & {\footnotesize {\em (0.28)}}
  \end{tabular}
        \caption{\label{tab:RunTime} Mean {\em (standard deviation)} run time per iteration of the sampling algorithm in seconds.}
\end{table}

To investigate the performance of the proposed procedure for conducting inference on band-collapsed measures as functions of outcome, we computed pointwise 95\% credible intervals for the nine band-collapsed curves $f^{HF}_{1}$, $f^{HF}_{2}$, $f^{HF}_{3}$, $f^{LF/HF}_{1}$, $f^{LF/HF}_{2}$, $f^{LF/HF}_{3}$, $\rho^{2, HF}_{12}$, $\rho^{2, HF}_{23}$, and $\rho^{2, HF}_{13}$. The mean and standard deviation of pointwise coverage probabilities integrated across $u$ are given in Table \ref{tab:CovBand}.   The integrated coverage was near the nominal 95\% level for each component, ranging between 94.3\%--97.4\%.  Coverage probabilities under different tuning parameters were indistinguishable.

\begin{table}
    \centering
\renewcommand\arraystretch{.75}
\begin{tabular}{l l l | c c c c c c c c c}
$n$&  $N$    &  $G$     & $f^{HF}_{1} $   &  $f^{HF}_{2}$       &  $f^{HF}_{3}$  &  $f^{LF/HF}_{1} $   &  $f^{LF/HF}_{2}$       &  $f^{LF/HF}_{3}$        &  $\rho^{2, HF}_{12}$  &  $\rho^{2, HF}_{23}$ & $\rho^{2, HF}_{13}$\\ \hline
300& 25 & $10^5$   &   .943   &  .968   &  .958     &  .969  &  .965 &  .961 &  .968  &  .951 &  .963 \\
&    &          &    {\footnotesize {\em (.112)}}   & {\footnotesize {\em (.073)}} & {\footnotesize \em (.097)} & {\footnotesize \em (.084)}& {\footnotesize \em (.078)} & {\footnotesize \em (.093)} & {\footnotesize \em (.058)}& {\footnotesize \em (.083)} & {\footnotesize \em (.078)} \\
 &&&&&&&&\\
   &    & $10^{10}$   &   .943   &  .968   &  .958     &  .969  &  .965 &  .961 &  .968  &  .951 &  .963 \\
&    &          &    {\footnotesize {\em (.112)}}   & {\footnotesize {\em (.073)}} & {\footnotesize \em (.097)} & {\footnotesize \em (.084)}& {\footnotesize \em (.078)} & {\footnotesize \em (.093)} & {\footnotesize \em (.058)}& {\footnotesize \em (.083)} & {\footnotesize \em (.078)} \\
 &&&&&&&&\\
500& 25 & $10^5$   &  .968&	.949&	.969&	.965&	.974&	.967&	.968&	.950&	.963 \\
&    &          &    {\footnotesize {\em (.070)}}   & {\footnotesize {\em (.087)}} & {\footnotesize \em (.070)} & {\footnotesize \em (.097)}& {\footnotesize \em (.062)} & {\footnotesize \em (.069)} & {\footnotesize \em (.068)}& {\footnotesize \em (.086)} & {\footnotesize \em (.083)}\\&&&&&&&&\\
   &    & $10^{10}$     &  .968&	.949&	.969&	.965&	.974&	.967&	.968&	.950&	.963 \\
&    &          &    {\footnotesize {\em (.070)}}   & {\footnotesize {\em (.087)}} & {\footnotesize \em (.070)} & {\footnotesize \em (.097)}& {\footnotesize \em (.062)} & {\footnotesize \em (.069)} & {\footnotesize \em (.068)}& {\footnotesize \em (.086)} & {\footnotesize \em (.083)}\\&&&&&&&&\\
300& 50 & $10^5$  &   .948   &  .961   &  .960     &  .962  &  .945 &  .945 &  .962  &  .961 &  .960 \\
&    &          &    {\footnotesize {\em (.107)}}   & {\footnotesize {\em (.077)}} & {\footnotesize \em (.073)} & {\footnotesize \em (.092)}& {\footnotesize \em (.110)} & {\footnotesize \em (.106)} & {\footnotesize \em (.082)}& {\footnotesize \em (.073)} & {\footnotesize \em (.081)}\\
 &&&&&&&&\\
   &    & $10^{10}$     &   .948   &  .961   &  .960     &  .962  &  .945 &  .945 &  .962  &  .961 &  .960 \\
&    &          &    {\footnotesize {\em (.107)}}   & {\footnotesize {\em (.077)}} & {\footnotesize \em (.073)} & {\footnotesize \em (.092)}& {\footnotesize \em (.110)} & {\footnotesize \em (.106)} & {\footnotesize \em (.082)}& {\footnotesize \em (.073)} & {\footnotesize \em (.081)}\\
 &&&&&&&&\\
500& 50 & $10^5$  &  .967 &	.956 &	.971 &	.954 &	.955 &	.963 &	.959 &	.953 & .969 \\
&    &          &    {\footnotesize {\em (.070)}}   & {\footnotesize {\em (.079)}} & {\footnotesize \em (.064)} & {\footnotesize \em (.102)}& {\footnotesize \em (.093)} & {\footnotesize \em (.078)} & {\footnotesize \em (.080)}& {\footnotesize \em (.088)} & {\footnotesize \em (.059)} \\
 &&&&&&&&\\
   &    & $10^5$  &  .967 &	.956 &	.971 &	.954 &	.955 &	.963 &	.959 &	.953 & .969 \\
&    &          &    {\footnotesize {\em (.070)}}   & {\footnotesize {\em (.079)}} & {\footnotesize \em (.064)} & {\footnotesize \em (.102)}& {\footnotesize \em (.093)} & {\footnotesize \em (.078)} & {\footnotesize \em (.080)}& {\footnotesize \em (.088)} & {\footnotesize \em (.059)} \\
\end{tabular}
        \caption{\label{tab:CovBand} Mean {\em (standard deviation)}  coverage of 95\% credible intervals for band-collapsed measures from 100 random samples of $N$ conditional MA(2) time series of length $n$ using hyperparamter $G$.}
\end{table}

To compare the performance of the proposed procedure to existing approaches, we also
computed two two-stage estimators of within-period band-collapsed measures.  In the first stage, periodograms were calculated and summed within HF and LF bands for each of the $P=3$ series for each of the $N$ subjects to obtain raw subject-specific estimates.  The raw subject-specific HF and LF/HF estimates were then smoothed across $u$.  For the first estimator, smoothing was achieved by fitting a cubic smoothing spline with smoothing parameter selected through generalized cross-validation (GCV) \citep{gu2013}.  For the second estimator, smoothing was achieved through local linear regression with plug-in bandwidth \citep{loader1999}.  We defined the integrated square error (ISE) of an estimate $\hat{f}^{HF}_p$ of $f^{HF}_p$ as
$$\int_0^1 \left[ \hat{f}^{HF}_p(u) - f^{HF}_p(u) \right]^2 du.$$
The ISEs for $\hat{f}^{LF/HF}_p$ and $\hat{\rho}^{2, \, HF}_{p q}$ were similarly defined.
 The mean and standard deviation of the ISEs are presented in Table \ref{tab:MSE}.  As expected, the ISE of each estimator improved with an increase in either $n$ or $N$.  The insensitivity of the proposed procedure to choice of hyperparameter that was observed through indistinguishable coverage probabilities was also observed in the ISE; the ISE under $G=10^5$ and $G=10^{10}$ were identical up to at least three significant digits.  In each setting, the proposed estimator had smaller mean ISE compared to the two-stage procedures.

   \begin{table}
    \centering
\renewcommand\arraystretch{.75}
\begin{tabular}{l l l | c c c c c c}
$n$&  $N$    &  Estimator     & $f^{HF}_{1}$  &  $f^{HF}_{2}$       &  $f^{HF}_{3}$  &  $f^{LF/HF}_{1} $   &  $f^{LF/HF}_{2}$     &  $f^{LF/HF}_{3}$ \\ \hline
300& 25 & Bayes: $10^5$   &  6.76 &	6.75 &	6.57 &	2.98 &	3.28 &	4.31 \\
&    &          &    {\footnotesize {\em (4.96)}}   & {\footnotesize {\em (6.03)}} & {\footnotesize \em (6.30)} & {\footnotesize \em (2.23)}& {\footnotesize \em (2.45)} & {\footnotesize \em (3.34)} \\
 &&&&&&&&\\
   &    & Bayes: $10^{10}$     &  6.76 &	6.75 &	6.57 &	2.98 &	3.28 &	4.31 \\
&    &          &    {\footnotesize {\em (4.96)}}   & {\footnotesize {\em (6.03)}} & {\footnotesize \em (6.30)} & {\footnotesize \em (2.23)}& {\footnotesize \em (2.45)} & {\footnotesize \em (3.34)} \\
 &&&&&&&&\\
  &    & 2-Stage: Spline   & 12.11 & 13.14 & 12.82 & 8.75 &	10.06 &	8.92 \\
&    &          &    {\footnotesize {\em (12.70)}}   & {\footnotesize {\em (18.56)}} & {\footnotesize \em (12.20)} & {\footnotesize \em (8.29)}& {\footnotesize \em (8.29)} & {\footnotesize \em (7.00)} \\
 &&&&&&&&\\
  &    & 2-Stage: LOESS   &   10.31 &	11.25 &	11.68 &	10.01 &	10.86 &	10.18 \\
&    &          &    {\footnotesize {\em (6.97)}}   & {\footnotesize {\em (14.06)}} & {\footnotesize \em (7.89)} & {\footnotesize \em (7.36)}& {\footnotesize \em (7.36)} & {\footnotesize \em (7.97)} \\
 &&&&&&&&\\
500& 25 & Bayes: $10^5$  & 3.75 &	4.57&	3.75&	1.86&	1.79&	2.62\\
&    &          &    {\footnotesize {\em (2.94)}}   & {\footnotesize {\em (3.54)}} & {\footnotesize \em (2.82)} & {\footnotesize \em (1.68)}& {\footnotesize \em (1.29)} & {\footnotesize \em (1.89)} \\
 &&&&&&&&\\
   &    & Bayes: $10^{10}$     & 3.75 &	4.57&	3.75&	1.86&	1.79&	2.62\\
&    &          &    {\footnotesize {\em (2.94)}}   & {\footnotesize {\em (3.54)}} & {\footnotesize \em (2.82)} & {\footnotesize \em (1.68)}& {\footnotesize \em (1.29)} & {\footnotesize \em (1.89)} \\
 &&&&&&&&\\
 &    & 2-Stage: Spline & 7.87&	8.05&	7.15&	5.26&	5.04&	4.60  \\
&    &          &    {\footnotesize {\em (7.76)}}   & {\footnotesize {\em (8.63)}} & {\footnotesize \em (7.27)} & {\footnotesize \em (4.35)}& {\footnotesize \em (4.80)} & {\footnotesize \em (3.44)} \\
 &&&&&&&&\\
  &    & 2-Stage: LOESS   &7.93&	7.69&	7.21&	5.94&	5.34&	5.42  \\
&    &          &    {\footnotesize {\em (6.32)}}   & {\footnotesize {\em (5.29)}} & {\footnotesize \em (5.61)} & {\footnotesize \em (3.84)}& {\footnotesize \em (3.50)} & {\footnotesize \em (2.87)} \\
 &&&&&&&&\\
300& 50 & Bayes: $10^5$  &  3.33 &	3.61	& 3.59	& 1.51 &	1.83	& 2.51  \\
&    &          &    {\footnotesize {\em (2.69)}}   & {\footnotesize {\em (2.84)}} & {\footnotesize \em (2.97)} & {\footnotesize \em (1.37)}& {\footnotesize \em (1.39)} & {\footnotesize \em (1.81)} \\
 &&&&&&&&\\
   &    & Bayes: $10^{10}$     &  3.33 &	3.61	& 3.59	& 1.51 &	1.83	& 2.51  \\
&    &          &    {\footnotesize {\em (2.69)}}   & {\footnotesize {\em (2.84)}} & {\footnotesize \em (2.97)} & {\footnotesize \em (1.37)}& {\footnotesize \em (1.39)} & {\footnotesize \em (1.81)} \\
 &&&&&&&&\\
 &    & 2-Stage: Spline   & 6.64 &	7.09	&6.84&	4.88&	5.39&	5.57  \\
&    &          &    {\footnotesize {\em (7.70)}}   & {\footnotesize {\em (10.34)}} & {\footnotesize \em (6.63)} & {\footnotesize \em (4.15)}& {\footnotesize \em (4.41)} & {\footnotesize \em (4.35)} \\
 &&&&&&&&\\
  &    & 2-Stage: LOESS   &   5.50 &	5.60 &	5.91 &	5.31 &	5.62 &	5.79 \\
&    &          &    {\footnotesize {\em (4.00)}}   & {\footnotesize {\em (6.35)}} & {\footnotesize \em (4.21)} & {\footnotesize \em (3.64)}& {\footnotesize \em (3.48)} & {\footnotesize \em (3.85)} \\
 &&&&&&&&\\
500& 50 & Bayes: $10^5$  & 1.74	& 2.12 &	2.06 & 1.00 &	1.13	& 1.44\\
&    &          &    {\footnotesize {\em (1.25)}}   & {\footnotesize {\em (1.50)}} & {\footnotesize \em (1.36)} & {\footnotesize \em (0.83)}& {\footnotesize \em (1.13)} & {\footnotesize \em (0.92)} \\
 &&&&&&&&\\
  &    & Bayes: $10^{10}$     & 1.74	& 2.12 &	2.06 & 1.00 &	1.13	& 1.44\\
&    &          &    {\footnotesize {\em (1.25)}}   & {\footnotesize {\em (1.50)}} & {\footnotesize \em (1.36)} & {\footnotesize \em (0.83)}& {\footnotesize \em (1.13)} & {\footnotesize \em (0.92)} \\
 &&&&&&&&\\
 &    & 2-Stage: Spline   & 4.52&	4.87&	3.84&	2.87&	3.19&	2.84  \\
&    &          &    {\footnotesize {\em (5.00)}}   & {\footnotesize {\em (5.31)}} & {\footnotesize \em (4.31)} & {\footnotesize \em (2.56)}& {\footnotesize \em (3.07)} & {\footnotesize \em (2.32)} \\
 &&&&&&&&\\
  &    & 2-Stage: LOESS   &  3.85&	3.89&	3.47&	2.81&	3.04&	2.79 \\
&    &          &    {\footnotesize {\em (3.13)}}   & {\footnotesize {\em (2.89)}} & {\footnotesize \em (2.41)} & {\footnotesize \em (1.92)}& {\footnotesize \em (2.18)} & {\footnotesize \em (1.57)} \\
\end{tabular}
        \caption{\label{tab:MSE} Mean {\em (standard deviation)} of the integrated square error (ISE) of band-collapsed measures from 100 random samples of $N$ independent conditional MA(2) time series of length $n$.  Estimates were obtained using the proposed procedure with tuning parameter $G=10^5$ (Bayes: $10^5$) and $G=10^{10}$ (Bayes: $10^{10}$) and two-stage estimators using smoothing splines (2-Stage: Spline) and local linear regression (2-Stage: LOESS).  Values are reported $\times 10^3$ for HF measures and $\times 10^{5}$ for LF/HF measures. }
\end{table}

\newpage

\section{Details of the Sampling Scheme}
\label{sec:samp_scheme}
In this appendix we provide more details about the sampling scheme outlined in
Section~4.4, assuming $P=3$. As in Section~4.4,
$\mb q'_{jm}$ is the row of
$Q$ corresponding to $u_j$ and $\omega_m$.
The DFT of the $p$th series from the $j$th subject at Fourier frequency $\omega_m$ is denoted by $Y_{pjm}$.
The $k \ell$ element of $\Theta(\omega_{m},u_{j})$ defined in Section~4.2 is expressed as
\begin{equation}
\theta_{k\ell jm}=\mb q'_{jm} \mb\eta_{rk\ell}+ i\, \mb q'_{jm}\mb\eta_{ik\ell},\;\; k > \ell=1,\ldots,P-1,
\label{eq:theta}
\end{equation}
where the $i$ in the second term on the right-hand side of (\ref{eq:theta}) is the
unit imaginary number.
The diagonal elements of $\Psi^{-1}(\omega,u)$ are expressed as
$
\psi_{kkjm}^{-1}=\exp(\mb q'_{jm}\mb\eta_{dkk}),\;\; k=1,\ldots,P.
$
To aid presentation and simplify notation, the superscript for iteration number is suppressed and all derived distributions are conditional on the current values of all other parameters.

\subsubsection*{Drawing the Basis Function Coefficient Vectors}
The conditional posterior distribution
of $\mb\eta_{ck\ell}$, $c=r,i$, $k>\ell=1,\ldots,P-1$, is
multivariate normal, $N(\mb\mu_{ck\ell},\Sigma_{ck\ell})$.
In what follows we provide expressions for $\mb\mu_{ck\ell}$ and $\Sigma_{ck\ell}$. \\
{\em Mean vectors and covariance matrices for $\mb\eta_{r21}$ and $\mb\eta_{i21}$}
\begin{eqnarray*}
\Sigma_{c21}^{-1} & = & 2\sum_{j=1}^N\sum_{m=1}^M \psi_{11jm}^{-1} |Y_{2jm}|^2 \mb q_{jm} \mb q'_{jm}+D_{c21}^{-1}
,\;\; c=r,i \\
\Sigma_{r21}^{-1}\mb \mu_{r21} &=& 2\sum_{j=1}^N\sum_{m=1}^M \psi_{11jm}^{-1}\Re\bigl\{Y_{1jm}Y_{2jm}^*-\theta_{31jm}^*Y_{2jm}Y_{3jm}^*\bigr\}\mb q_{jm} \\
\Sigma_{i21}^{-1}\mb \mu_{i21} &=& 2\sum_{j=1}^N\sum_{m=1}^M \psi_{11jm}^{-1}\Im\bigl\{Y_{1jm}Y_{2jm}^*+
\theta_{31jm}^*Y_{2jm}Y_{3jm}^*\bigr\}\mb q_{jm}. \\
\end{eqnarray*}
Note that $\psi_{11jm}^{-1}$ and $\theta_{31jm}$ are evaluated at their current values. \\
{\em Mean vectors and covariance matrices for $\mb\eta_{r31}$ and $\mb\eta_{i31}$}
\begin{eqnarray*}
\Sigma_{c31}^{-1} & = & 2\sum_{j=1}^N\sum_{m=1}^M \psi_{11jm}^{-1} |Y_{3jm}|^2 \mb q_{jm} \mb q'_{jm}+D_{c31}^{-1}
,\;\; c=r,i \\
\Sigma_{r31}^{-1}\mb \mu_{r31} &=& 2\sum_{j=1}^N\sum_{m=1}^M \psi_{11jm}^{-1}\Re\bigl\{Y_{1jm}Y_{3jm}^*-\theta_{21jm}^*Y_{2jm}^*Y_{3jm}\bigr\}\mb q_{jm} \\
\Sigma_{i31}^{-1}\mb \mu_{i31} &=& 2\sum_{j=1}^N\sum_{m=1}^M \psi_{11jm}^{-1}\Im\bigl\{Y_{1jm}Y_{3jm}^*+
\theta_{21jm}^* Y_{2jm}^*Y_{3jm}\bigr\}\mb q_{jm}. \\
\end{eqnarray*}
{\em Mean vectors and covariance matrices for $\mb\eta_{r32}$ and $\mb\eta_{i32}$}
\begin{eqnarray*}
\Sigma_{c32}^{-1} & = & 2\sum_{j=1}^N\sum_{m=1}^M \psi_{22jm}^{-1} |Y_{3jm}|^2 \mb q_{jm} \mb q'_{jm}+D_{c32}^{-1}
,\;\; c=r,i \\
\Sigma_{r32}^{-1}\mb \mu_{r32} &=& 2\sum_{j=1}^N\sum_{m=1}^M \psi_{22jm}^{-1}\Re\bigl\{Y_{2jm}^*Y_{3jm}\bigr\}
\mb q_{jm} \\
\Sigma_{i32}^{-1}\mb \mu_{i32} &=& 2\sum_{j=1}^N\sum_{m=1}^M \psi_{22jm}^{-1}\Im\bigl\{Y_{2jm}Y_{3jm}^*
\bigr\}\mb q_{jm}. \\
\end{eqnarray*}
The basis function coefficient vectors $\mb\eta_{dkk}$, $k=1,\dots,P$, are drawn from
$p\bigl(\mb\eta_{dkk} \mid Q,\mb v_k, D_{dkk}\bigr)$, given in Section~4.3.
The entries $v_{kjm}$ of $\mb v_k$ for $k=1,2,3$ are as follows.
\begin{eqnarray*}
v_{1jm} & = & |Y_{1jm}|^2+|\theta_{21jm}Y_{2jm}|^2+|\theta_{31jm}Y_{3jm}|^2 \\
& - & 2\Re\bigl\{\theta_{21jm}Y_{1jm}^*Y_{2jm}+\theta_{31jm}Y_{1jm}^*Y_{3jm}
-\theta_{21jm}^*\theta_{31jm}Y_{2jm}^* Y_{3jm}\bigr\}. \\
v_{2jm} & = & |Y_{2jm}|^2+|\theta_{32jm}Y_{3jm}|^2-2\Re\bigl\{\theta_{32jm}Y_{2jm}^*Y_{3jm} \bigr\}. \\
v_{3jm} & = & |Y_{3jm}|^2.
\end{eqnarray*}
The vectors $\mb\eta_{dkk}$, $k=1,\ldots,P$, are generated independently via a Metropolis-Hastings step with
a multivariate $t$ proposal distribution, $t_\nu(\mb{\hat{\eta}}_{dkk},\hat{\Sigma}_{dkk})$, where
\[
\mb{\hat{\eta}}_{dkk}=\argmax{\mb\eta_{dkk}}\log p\bigl(\mb\eta_{dkk} \mid Q,\mb v_k, D_{dkk}\bigr)
\]
and
\[
\hat{\Sigma}_{dkk}=\Bigl[-\frac{\partial^2}{\partial\mb\eta_{dkk}\partial\mb\eta'_{dkk}}\log p\bigl(\mb\eta_{dkk} \mid Q,\mb v_k, D_{dkk}\bigr)  \Bigr]_{\mb\eta_{dkk}=\mb{\hat{\eta}}_{dkk}}^{-1}.
\]
The gradient and Hessian of $\log p\bigl(\mb\eta_{dkk} \mid Q,\mb v_k, D_{dkk}\bigr)$ are given by
\[
\sum_{j=1}^N\sum_{m=1}^M \Bigl[1-v_{kjm}\exp(\mb q'_{jm}\mb\eta_{dkk})\Bigr]\mb q_{jm}-D_{dkk}^{-1}\mb\eta_{dkk}
\]
and
\[
-\sum_{j=1}^N\sum_{m=1}^M v_{kjm}\exp(\mb q'_{jm}\mb\eta_{dkk}) \mb q_{jm} \mb q'_{jm} -D_{dkk}^{-1},
\]
respectively.
\subsubsection*{Drawing the Smoothing Parameters}
Details are given below for the smoothing parameters associated with the real part
of $\theta_{k\ell}(\omega,u)$. The details for the rest of the smoothing parameters are similar. The smoothing parameters $\tau_{\beta rk\ell}^2$, $\tau_{\gamma rk\ell}^2$ and
$\tau_{\delta rk\ell}^2$ are drawn independently for $k>\ell=1,\ldots,P-1$, as follows.
\begin{eqnarray*}
\tau_{\beta rkl}^2 & \ind & IG((n_b+\nu)/2,\,\mb b'_{rkl} \mb b_{rkl}/2 +\nu/g_{\beta rkl}) \\
\tau_{\gamma rkl}^2 & \ind & IG((n_c+\nu)/2,\,\mb c'_{rkl} \mb c_{rkl}/2 +\nu/g_{\gamma rkl})
\\
\tau_{\delta rkl}^2 & \ind & IG((n_d+\nu)/2,\,\mb d'_{rkl} \mb d_{rkl}/2 +\nu/g_{\delta rkl}),
\end{eqnarray*}
where
\begin{eqnarray*}
g_{\beta rkl} & \ind & IG((\nu+1)/2,\, \nu/\tau_{\beta rkl}^2+1/G^2) \\
g_{\gamma rkl} & \ind & IG((\nu+1)/2,\, \nu/\tau_{\gamma rkl}^2+1/G^2) \\
g_{\delta rkl} & \ind & IG((\nu+1)/2,\, \nu/\tau_{\delta rkl}^2+1/G^2). \\
\end{eqnarray*}

\end{document}